\documentclass[
reprint,
nofootinbib,
amsmath,
amssymb,
aps,
]{revtex4-2}
\usepackage{ulem}
\usepackage{graphicx}
\usepackage{bm}
\usepackage{hyperref}
\usepackage[all]{hypcap}
\usepackage{xspace}
\usepackage[usenames,dvipsnames,svgnames]{xcolor}
\usepackage{booktabs}
\definecolor{darkgreen}{cmyk}{0.9,0,0.9,0.6}
\definecolor{brickred}{RGB}{210,65,84}
\hypersetup{colorlinks=true,
citecolor=Blue,
linkcolor=Magenta,
filecolor=NavyBlue,
urlcolor=NavyBlue}

\graphicspath{{figure/}}

\newcommand{\rhoint}{{\rho_{\rm int}}}

\begin{document}

\title{Sound of Keplerian binaries in a homogeneous Newtonian fluid background}

\author{Yan Cao$^{a}$}
\email[Contact author: ]{yancao@smail.nju.edu.cn}

\author{Mu-Chun Chen$^{b,c}$}

\affiliation{\begin{footnotesize}
${}^a$School of Physics, Nanjing University, Nanjing 210093, China\\
${}^b$	Tsung-Dao Lee Institute (TDLI),
No.1 Lisuo Road, 201210 Shanghai, China\\
${}^c$ 	School of Physics and Astronomy, Shanghai Jiao Tong University,
800 Dongchuan Road, 200240 Shanghai, China
\end{footnotesize}}

\begin{abstract}
We formulate the dissipation due to sound-wave emission from a Keplerian binary embedded in a homogeneous Newtonian fluid background at rest in the binary center-of-mass frame within linear perturbation theory. Rather than starting from the local wake force, we treat the binary as a localized periodic mass source and compute the sound-wave fluxes carried away by the medium. The time-averaged energy, linear-momentum, and angular-momentum fluxes are obtained from Noether currents of the fluid perturbations. For a classical inviscid barotropic Newtonian fluid, retaining the self-gravity of density perturbations while adopting the Jeans prescription, we derive explicit flux formulas for circular and eccentric binaries with arbitrary mass ratio. This formulation also clarifies the ultraviolet behavior of point sources: in a nondispersive fluid, the harmonic sum becomes cutoff-sensitive precisely when at least one component becomes supersonic. We examine explicitly how finite source size removes this divergence using a uniform-sphere model and relate the resulting regularization to a corresponding effective harmonic cutoff. The associated secular eccentricity evolution exhibits circularization in the subsonic regime when the fluid self-gravity is sufficiently weak but can change sign outside this regime, with eccentricity growth in the strictly supersonic UV limit. For a Schr{\"o}dinger-Poisson quantum fluid, quantum pressure instead softens the high-frequency contribution and renders the point-source fluxes UV finite within the nonrelativistic regime; our numerical results further indicate eccentricity growth due to sound radiation over the parameter space explored. Finally, we translate orbital-decay measurements of two binary pulsars into order-of-magnitude bounds on the homogeneous quantum-fluid density surrounding each system.
\end{abstract}

\maketitle

\tableofcontents

\onecolumngrid

\section{Introduction}
Binary systems embedded in astrophysical environments can lose orbital energy and angular momentum by gravitationally exciting density perturbations in the surrounding medium. This environmental dissipation, known as dynamical friction, arises because each moving body excites a wake whose gravitational field reacts back on the source. The theory of dynamical friction originates in the classic collisionless and gaseous analyses~\cite{Chandrasekhar:1943ys,Ostriker_1999} and has since been developed for circular motion, eccentric Keplerian motion, and binary systems in classical fluids~\cite{Kim_2007,Kim_2008,Desjacques_2022,Buehler_2024,2024ApJ...974..216O,2026ApJ..1000..212O,Eytan:2025eccentric}. Similar effects have also been studied in superfluids and fuzzy or ultralight-dark-matter media, where the gravitational response can be formulated either in terms of a classical wave field or in its equivalent quantum-fluid description~\cite{Hui:2016ltb,Berezhiani_2019,Annulli:2020ultralight,Buehler:2022fdm,Berezhiani:2023vlo,Koo:2025jkx,Wang_2022,Lancaster_2020,Glennon_2024,boey2024dynamicalfrictionblackholes,szpilfidel2025fuzzydarkmatterdynamical}. Recent work has further examined supermassive-black-hole binaries in more realistic ultralight-dark-matter halo environments~\cite{Koo:2023,Bromley:2023}.

Most treatments are organized around the local wake force. For a prescribed periodic source in linear response, however, the same dissipative physics can be formulated as a radiation problem. In this paper, we consider a Keplerian binary in an infinite homogeneous Newtonian medium at rest in the binary center-of-mass frame and assume that the medium-induced backreaction changes the prescribed orbit only secularly. The self-gravity of density perturbations is retained, while the gravitational field of the unperturbed homogeneous background is subtracted through the usual Jeans prescription (the ``Jeans swindle''). We set the free modes to zero and retain only the outgoing periodically forced response. Treating the binary as a localized periodic mass source, we solve for the outgoing sound waves in the frequency domain and compute the energy, linear-momentum, and angular-momentum fluxes carried into the far zone, following the same strategy used in treatments of massive-field radiation reaction~\cite{LI198620,PhysRevD.49.6892,Cheng:2023qys,Chen:2025mwl}. This viewpoint complements the wake-force description by placing the dissipative fluxes in a common framework. It also separates the kinematics of the periodic source from the dispersion relation and mode normalization of the medium, making the comparison between the classical- and quantum-fluid cases particularly transparent.

Starting from the Lagrangian of a classical inviscid barotropic fluid, we derive the far-zone wave amplitude and the associated Noether fluxes for an arbitrary localized periodic source, and then specialize the results to circular and eccentric binaries of arbitrary mass ratio. Our results for the time-averaged energy and angular-momentum fluxes agree with the time-averaged dissipated power and torque obtained from direct force calculations in Refs.~\cite{Desjacques_2022,Eytan:2025eccentric}. The radiation formulation directly expresses the dissipative fluxes as sums over the orbital harmonics, making their UV behavior particularly transparent. A stationary-phase analysis then shows that, for the classical fluid considered here, the harmonic sums for point sources become logarithmically sensitive to the UV cutoff precisely when at least one binary component becomes supersonic somewhere along the orbit.

We then repeat the construction for a Schr{\"o}dinger-Poisson quantum fluid. Quantum pressure modifies the high-frequency dispersion relation, rendering the point-source fluxes UV convergent within the nonrelativistic regime. We derive the fluxes for circular and eccentric binaries and their quadrupole-order approximations. For circular binaries, we also obtain the large-$\mathcal{M}_Q$ asymptotic behavior of the energy flux. Finally, as an illustration, we use binary-pulsar orbital decay to place order-of-magnitude upper bounds on the densities of homogeneous quantum-fluid backgrounds surrounding these systems. The numerical results in this paper were obtained using publicly available code.\footnote{The code used to generate the numerical results is publicly available at \url{https://github.com/cao-yan-phys/binary-sound-wave-flux-calculators}.}

This paper is organized as follows. In Sec.~\ref{sec:1}, we formulate the radiation problem for a classical inviscid barotropic fluid, derive the Noether fluxes, and apply the results to circular and eccentric Keplerian binaries, followed by analyses of the quadrupole-order approximation and UV behavior. In Sec.~\ref{sec:quantum}, we repeat the construction for a Schr{\"o}dinger-Poisson quantum fluid. In Sec.~\ref{sec:constraints}, we use binary-pulsar orbital decay measurements to obtain illustrative bounds on homogeneous quantum-fluid background densities. Table~\ref{tab:notation} summarizes the notation used in the classical- and quantum-fluid formulations. Throughout this paper, we use natural units with $G=c=\hbar=1$.

\section{Binary in a classical fluid}\label{sec:1}
We consider a classical inviscid Newtonian fluid in a flat spacetime. The fluid is described by its mass density $\rho$, velocity $\mathbf{v}$, pressure $P$, and Newtonian potential $\varPhi$, which satisfy the equations of motion
\begin{equation}
\begin{aligned}
\dot\rho + \nabla\cdot(\rho\mathbf{v})&=0,
\\
\dot{\mathbf{v}}+\mathbf{v}\cdot\nabla \mathbf{v}
&=-\frac{\nabla P}{\rho}-\nabla\varPhi,
\\
\nabla^2\varPhi &=4\pi(\rho-\bar\rho+\rho_*)
, \label{classical_fluid}
\end{aligned}
\end{equation}
where $\rho_*$ denotes the mass density external to the fluid, and the gravitational field associated with the homogeneous background $\bar\rho$ is removed by the usual Jeans prescription, corresponding to $\rho\to\rho-\bar\rho$ in the Poisson equation. For $\rho_*=0$, a static homogeneous background solution to \eqref{classical_fluid} is given by $\nabla\bar\rho=\nabla\bar P=\nabla\bar\varPhi=\bar{\mathbf{v}}=\mathbf{0}$. We are interested in the linear perturbation of this fluid background by a set of nonrelativistic point masses, corresponding to the perturber density $\rho_*=\sum_Im_I\,\delta^3(\mathbf{x}-\mathbf{X}_I(t))$, where $\mathbf{X}_I(t)$ denotes the coordinates of the $I$th body with mass $m_I$. We decompose the fields as
\begin{equation}
\rho=\bar\rho(1+\alpha), \quad P=\bar P + p, \quad \varPhi=\bar\varPhi + \Phi, \quad \mathbf{v}=\bar{\mathbf{v}}+\mathbf{u}.
\end{equation}
The linearized fluid equations then read
\begin{equation} \label{system}
\begin{aligned}
\nabla^2\Phi &=
4\pi (\bar\rho\,\alpha+\rho_*),
\\
\dot{\mathbf{u}} &=-\frac{\nabla p}{\bar\rho} -\nabla\Phi
,
\\
\dot\alpha+\nabla\cdot\mathbf{u} &=0.
\end{aligned}
\end{equation}
We assume that the fluid is barotropic, with squared sound speed $c_s^2\equiv p/(\bar\rho\,\alpha)$. We refer to this model below as the ``classical fluid.'' It follows that
\begin{equation}
\ddot\alpha-c_s^2\nabla^2\alpha=4\pi (\bar\rho\, \alpha+\rho_*). \label{sound_wave_eq}
\end{equation}
A time-dependent source $\rho_*(t,\mathbf{x})$ thus generates sound waves described by the density perturbation $\alpha$. Note that the first term on the right-hand side of Eq.~\eqref{sound_wave_eq} arises from the self-gravity of the fluid.

We now decompose the velocity perturbation as $\mathbf{u}=\nabla s +\mathbf{U}$, with $\nabla\cdot\mathbf{U}=0$. The equations become
\begin{equation}
\begin{aligned}
\nabla^2\Phi &=
4\pi (\bar\rho\alpha+\rho_*),
\\
\nabla \dot s +\dot{\mathbf{U}} &=-c_s^2\nabla\alpha-\nabla\Phi
,
\\
\dot\alpha+\nabla^2s &=0.
\end{aligned}\label{linearized}
\end{equation}
We can write $\dot{\mathbf{U}}=\nabla\chi$. Since $\nabla\cdot\mathbf{U}=0$, it follows that $\nabla\cdot\dot{\mathbf{U}}=0$, and therefore $\nabla^2\chi=0$. Assuming that $\chi$ vanishes, or approaches a constant, at spatial infinity, the uniqueness theorem for Laplace's equation implies that $\chi=0$ up to an irrelevant constant. Therefore, $\dot{\mathbf{U}}=0$, and $\dot s = -c_s^2\alpha-\Phi+C(t)$. Since $C(t)$ can be absorbed into a redefinition of $s$, we choose the gauge:
\begin{equation}
\dot{s}=-c_s^2\alpha-\Phi . \label{Bernoulli}
\end{equation}
The variable $s$ is the velocity potential for the irrotational velocity perturbation, complementary to the density perturbation $\alpha$, and now satisfies the linearized Bernoulli relation \eqref{Bernoulli}.

\subsection{Noether currents}
We proceed to analyze some consequences of Eq.~\eqref{system}, with $c_s^2= p/(\bar\rho\,\alpha)$. Defining $\nabla^2\Phi_*=4\pi \rho_*$ and $\nabla^2\Phi_\text{m}=4\pi \bar\rho\,\alpha$, we obtain from Eq.~\eqref{system} that
\begin{equation}\label{conservation_law_1}
\partial_t\left(\frac{1}{2}\bar\rho  u^2+\frac{1}{2}c_s^2\bar\rho\,\alpha^2+\frac{1}{8\pi}|\nabla\Phi_\text{m}|^2+\bar\rho\,\alpha\Phi_\text{m}\right)+\nabla\cdot\left(p\mathbf{u}-\frac{1}{4\pi}\dot\Phi_\text{m}\,\nabla\Phi_\text{m}+\bar\rho \Phi_\text{m}\mathbf{u}\right)=-\bar\rho \mathbf{u}\cdot\nabla\Phi_*,
\end{equation}
we can thus identify
\begin{equation}
\mathcal{E}=\frac{1}{2}\bar\rho  u^2+\frac{1}{2}c_s^2\bar\rho\,\alpha^2+\frac{1}{8\pi}|\nabla\Phi_\text{m}|^2+\bar\rho\,\alpha\Phi_\text{m} \label{energy_density}
\end{equation}
as the energy density, and
\begin{equation}
\mathbf{S}=p\mathbf{u}-\frac{1}{4\pi}\dot\Phi_\text{m}\,\nabla\Phi_\text{m}+\bar\rho \Phi_\text{m}\mathbf{u} \label{energy_flux_density}
\end{equation}
as the energy flux density. It should be stressed that this choice is not unique; the total radiation flux considered below, however, is independent of such choices. For $\rho_*=\sum_Im_I\,\delta^3(\mathbf{x}-\mathbf{X}_I(t))$, the right-hand side of Eq.~\eqref{conservation_law_1} contains the power delivered by the external gravitational field of the point masses to the fluid perturbation, since
\begin{equation}
\int d^3x\,\left(-\bar\rho \,\mathbf{u}\cdot\nabla\Phi_*\right)
=
-\sum_I m_I\mathbf{v}_I\cdot\left[\int d^3x\,\bar\rho \,\alpha \frac{\mathbf{x}-\mathbf{X}_I}{|\mathbf{x}-\mathbf{X}_I|^3}\right]+\frac{d}{dt}\left[
\sum_I m_I\int d^3x\,\frac{\bar\rho\alpha}{|\mathbf{x}-\mathbf{X}_I|}
\right].
\end{equation}

We assume that the source is prescribed and periodic, with period $T$. This approximation is appropriate when the fluid perturbation is weak and on timescales short compared with the backreaction time. We also set the free (homogeneous) modes of the $\alpha$ field to zero. The outgoing sound-wave flux is therefore supplied entirely by the source, yielding the corresponding energy-balance relation.

To simplify the analysis, we now turn to a Lagrangian formulation,\footnote{In view of Eq.~\eqref{sound_wave_eq}, one might try to construct an effective phonon Lagrangian $\mathcal{L}_\alpha=A\left[\frac{1}{2}(\dot\alpha)^2-\frac{1}{2}c_s^2|\nabla\alpha|^2+\frac{1}{2}(4\pi \bar\rho)\alpha^2+4\pi\rho_*\alpha\right]$. Its Noether currents, however, do not give the physical radiation fluxes with a mode-independent normalization $A$; see Sec.~\ref{sec:common_fluxes} for a discussion.} which makes the derivation of the radiation fluxes more convenient. The linearized fluid equations \eqref{linearized} can be obtained from the following Lagrangian for the perturbation variables $\{s,\alpha,\Phi\}$
\begin{equation}
\mathcal{L}=-\frac{\bar\rho }{2}|\nabla s|^2-\frac{\bar\rho }{2} c_s^2\alpha^2 - \bar\rho \Phi\alpha + \bar\rho s\dot \alpha-\frac{1}{8\pi}|\nabla\Phi|^2-\rho_*\Phi.
\end{equation}
From Noether's theorem (setting $\rho_*=0$), we obtain the energy flux density
\begin{equation}
\begin{aligned}
\mathbf{S} &=\frac{\partial\mathcal{L}}{\partial \nabla s}\dot s+\frac{\partial\mathcal{L}}{\partial \nabla \Phi}\dot \Phi
=-\bar\rho  \dot s\nabla s-\frac{1}{4\pi}\dot\Phi\nabla\Phi
\end{aligned}
\end{equation}
and the energy density
\begin{equation}
\begin{aligned}
\mathcal{E}
&=\frac{\partial\mathcal{L}}{\partial \dot \alpha}\dot \alpha-\mathcal{L}
=\frac{\bar\rho }{2}|\nabla s|^2 + \frac{1}{8\pi}|\nabla\Phi|^2 + \frac{\bar\rho }{2}c_s^2\alpha^2 + \bar\rho  \Phi \alpha
.
\end{aligned}
\end{equation}
These expressions reproduce Eqs.~\eqref{energy_density}--\eqref{energy_flux_density}. The linear momentum flux density is
\begin{equation}
\begin{aligned}
\Pi_q^i &=-\left[\frac{\partial \mathcal{L}}{\partial (\partial_i s)}\partial_qs
+\frac{\partial \mathcal{L}}{\partial (\partial_i\Phi)}\partial_q\Phi
-\delta_{iq}\mathcal{L}\right]
=\bar\rho \partial_i s \partial_q s
+\frac{1}{4\pi}\partial_i\Phi \partial_q\Phi
+\delta_{iq} \mathcal{L}
,
\end{aligned}
\end{equation}
the linear momentum density is
\begin{equation}
\begin{aligned}
p_q &=-\frac{\partial \mathcal{L}}{\partial \dot\alpha}\partial_q\alpha
=-\bar\rho s \partial_q \alpha
,
\end{aligned}
\end{equation}
the angular momentum flux density about the origin is\footnote{Henceforth, all angular momenta considered below are taken about the origin.}
\begin{equation}
\begin{aligned}
j_q^i &=\frac{1}{2}\epsilon_{qkl}\left[
\frac{\partial \mathcal{L}}{\partial (\partial_i s)}A^j _{kl}\partial_j s +\frac{\partial \mathcal{L}}{\partial (\partial_i \Phi)}A^j _{kl}\partial_j \Phi -A^i _{kl}\mathcal{L}
\right]
\\
&=\epsilon_{qkl}x^l\left(
-\bar\rho \partial_i s \partial_k s-\frac{1}{4\pi}\partial_i \Phi \partial_k \Phi
-\delta_{ik}\mathcal{L}
\right)
,
\end{aligned}
\end{equation}
and the angular momentum density is
\begin{equation}
\begin{aligned}
j_q & = \frac{1}{2}\epsilon_{qkl}\left(
\frac{\partial \mathcal{L}}{\partial \dot \alpha}A^j _{kl}\partial_j \alpha
\right)
= \bar\rho\epsilon_{qkl} x^l s\partial_k \alpha.
\end{aligned}
\end{equation}
These local expressions correspond to a particular choice of Noether currents.

\subsection{Radiation from a localized periodic source}
We proceed to evaluate the time-averaged radiation fluxes for a localized periodic sound source. To this end, we need a general expression for the sound waves. For convenience, we introduce an auxiliary time coordinate $\mathtt{t}\equiv c_st$ so that Eq.~\eqref{sound_wave_eq} becomes
\begin{equation}
\partial_{\mathtt t}^2\alpha-\nabla^2\alpha-m^2\alpha=n(\mathtt{t},\mathbf{x})\equiv\frac{4\pi}{c_s^2} \,\rho_*(t=\mathtt{t}/c_s,\mathbf{x}), \label{redefined_sound}
\end{equation}
where the effective tachyonic mass (or the Jeans wave number) is defined by
\begin{equation}
m\equiv \frac{\sqrt{4\pi \bar\rho} }{c_s}.
\end{equation}
The dispersion relation of a free plane wave $\alpha\propto e^{i(\mathbf{k\cdot x}-\omega \mathtt{t})}$ is $\omega^2=|\mathbf{k}|^2-m^2$, which exhibits a Jeans instability for $|\mathbf{k}|<m$. This instability, however, does not affect the forced radiative solution; the source fixes the wave frequency, for which $|\mathbf{k}|=\sqrt{\omega^2+m^2}$. The free modes are set to zero by the imposed boundary conditions, leaving only the forced linear response of the medium.

In the frequency domain, Eq.~\eqref{redefined_sound} reads $(\nabla^2+m^2+\omega^2)\tilde\alpha=-\tilde n$, with
\begin{equation}
\tilde \alpha(\omega,\mathbf{x})=\int_{-\infty}^\infty d\mathtt{t}\,\alpha(\mathtt{t},\mathbf{x})\,e^{i\omega \mathtt{t}},
\quad
\tilde n(\omega,\mathbf{x})=\int_{-\infty}^\infty d\mathtt{t}\,n(\mathtt{t},\mathbf{x})\,e^{i\omega \mathtt{t}}.
\end{equation}
For $\omega>0$, the outgoing solution to $(\nabla^2 + m^2+\omega^2) f(\mathbf{x},\mathbf{x}')=\delta^3(\mathbf{x}-\mathbf{x}')$ is
\begin{equation}\label{Green}
f(\mathbf{x},\mathbf{x}')=\int\frac{d^3k}{(2\pi)^3} \frac{e^{i\mathbf{k\cdot (x-x')}}}{-|\mathbf{k}|^2+m^2+(\omega+i\,0^+)^2}=-\frac{e^{i\sqrt{\omega^2+m^2}|\mathbf{x-x'}|}}{4\pi|\mathbf{x-x'}|}.
\end{equation}
The negative-frequency branch in the two-sided representation is obtained by complex conjugation. Taking the wave-zone limit of this retarded Green's function, the radiation field for a source with period $T\equiv2\pi/\tilde\Omega$ is\footnote{Nonperiodic sources require a different treatment, as the source evolution is no longer characterized by a discrete set of harmonics. See Ref.~\cite{2026ApJ..1000..212O} for a study of hyperbolic binary scattering in a gaseous background.}~\cite{Chen:2025mwl}
\begin{align}
\alpha(\mathtt{t},\mathbf{x}=R\,\mathbf{n})&=\frac{1}{4\pi R}\sum_{n}n_{(n)}\,e^{i\left[\mathbf{k}^{(n)}\cdot\mathbf{x}-\Omega_n \mathtt{t}\right]}, \label{radiation_1}
\\
n_{(n)} & = \int d^3x\,e^{-i\mathbf{k}^{(n)}\cdot\mathbf{x}}\left[\frac{1}{\mathtt{T}}\int_0^\mathtt{T}d\mathtt{t}\,n(\mathtt{t},\mathbf{x})\,e^{i\Omega_n \mathtt{t}}\right], \label{source}
\end{align}
where $\mathtt{T}\equiv 2\pi/\Omega=c_sT$, $\Omega_n\equiv n\Omega=n\tilde \Omega/c_s$, $\mathbf{k}^{(n)}=k_n\mathbf{n}=\Omega_n\sqrt{1+(n_0/n)^2}\,\mathbf{n}$ (note that $k_{-n}=-k_n$), $|\mathbf{n}|=1$, and $n_0\equiv m/\Omega=\sqrt{4\pi\bar\rho}\,/\tilde\Omega$.\footnote{For the Keplerian binary considered below, with $\tilde\Omega^2=M/a^3$, $n_0$ measures the importance of fluid self-gravity on the orbital timescale. The characteristic Jeans instability timescale is $t_J=1/(n_0\tilde\Omega)$, while the background fluid mass within a sphere of radius $a$ is $(4\pi/3)a^3\bar\rho=n_0^2M/3$. Thus, the regime $n_0\gtrsim1$ should be regarded as increasingly idealized for a static homogeneous background surrounding a prescribed Keplerian binary.
} See Appendix~\ref{appendix:1} for a time-domain derivation. Using the local plane-wave relation $-i\nabla \leftrightarrow \mathbf{k}$ for the radiation field, we also have
\begin{equation}
\begin{aligned} \label{radiation_2}
\left(\begin{matrix}
\Phi \\ s
\end{matrix}\right)&=\frac{1}{4\pi R}\sum_n\left(\begin{matrix}
\kappa_n^\Phi\\ \kappa_n^s
\end{matrix}\right) n_{(n)} \,e^{i\left[\mathbf{k}^{(n)}\cdot\mathbf{x}-\Omega_n \mathtt{t}\right]},
\end{aligned}
\end{equation}
with
\begin{equation}
\kappa^\Phi_n = \frac{-4\pi\bar\rho}{k_n^2},
\quad
\kappa^s_n = \frac{-i c_s\Omega_n}{k_n^2}.
\end{equation}

Using Eqs.~\eqref{radiation_1} and~\eqref{radiation_2}, the time-averaged energy flux is
\begin{equation}
\begin{aligned}\label{energy_flux}
P & = \left\langle R^2\int d\Omega_\mathbf{n}\,\mathbf{S}\cdot \mathbf{n} \right\rangle
=\int d\Omega_\mathbf{n}\sum_{n\ge 1}\frac{\bar\rho c_s^3}{8\pi^2}\frac{\Omega_n|n_{(n)}|^2}{k_n}
.
\end{aligned}
\end{equation}
Here, $d\Omega_\mathbf{n}$ denotes the element of solid angle, and $\langle f\rangle \equiv \int_0^{\mathtt T}\frac{d\mathtt t}{\mathtt T}\,f(\mathtt t) = \int_0^T\frac{dt}{T}\,f(t)$ denotes the time average over one source period. Henceforth, we consider time-averaged quantities unless stated otherwise. The time-averaged energy density is given by
\begin{equation}
\begin{aligned}
\mathcal{E}
&= \left\langle\frac{\bar\rho }{2}|\nabla s|^2 + \frac{1}{8\pi}|\nabla\Phi|^2 + \frac{\bar\rho }{2}c_s^2\alpha^2 + \bar\rho  \Phi \alpha \right\rangle
= \frac{1}{R^2}\sum_{n\ge 1}\frac{\bar\rho c_s^2 \Omega_n^2}{8\pi^2k_n^2}|n_{(n)}|^2.
\end{aligned}
\end{equation}
As expected, this quantity satisfies $\mathbf{S}\cdot\mathbf n=\mathcal{E}v_g$ mode by mode, where $v_g=c_sk_n/\Omega_n$ is the group velocity of the sound wave. The time-averaged linear momentum density is
\begin{equation}
\begin{aligned}
p_q & = \left\langle -\bar\rho s \partial_q \alpha \right\rangle=\sum_{n\ge 1} \frac{\bar\rho c_s \Omega_n}{8\pi^2k_n R^2}|n_{(n)}|^2n^q
,
\end{aligned}
\end{equation}
so the time-averaged linear momentum flux is
\begin{equation}\label{linear_momentum_flux}
\begin{aligned}
\mathcal{F}_q & = \sum_{n\ge 1}\int d\Omega_\mathbf{n}\, \frac{\bar\rho c_s^2 }{8\pi^2}|n_{(n)}|^2n^q
,
\end{aligned}
\end{equation}
which is consistent with the direct evaluation: $\mathcal{F}_q = \left\langle R^2\int d\Omega_\mathbf{n}\,\Pi_q^i n^i \right\rangle$. The time-averaged angular momentum density is
\begin{equation}
\begin{aligned}
j_q & = \left\langle \bar\rho\epsilon_{qkl} x^l s\partial_k \alpha \right\rangle
=\sum_n(-i)\frac{\bar\rho c_s \Omega_n}{16\pi^2 k_n^2 R^2}\epsilon_{qkl}x^ln_{(n)}\partial_k n_{(n)}^*
.
\end{aligned}
\end{equation}
Using $n_{(-n)}(\mathbf n)=n_{(n)}^*(\mathbf n)$, the time-averaged angular momentum flux is
\begin{equation}
\begin{aligned}\label{angular_momentum_flux}
\tau_q & = \int d\Omega_\mathbf{n}\sum_{n\neq0}(-i)\frac{\bar\rho c_s^2}{16\pi^2 k_n}\epsilon_{qkl}x^ln_{(n)}\partial_k n_{(n)}^*
\\
&=
\operatorname{Re}\int d\Omega_\mathbf n\sum_{n\ge1}(-i)\frac{\bar\rho c_s^2}{8\pi^2 k_n}\epsilon_{qkl}x^l n_{(n)}\partial_k n_{(n)}^*
.
\end{aligned}
\end{equation}
This result is consistent with the direct evaluation, $\tau_q = \left\langle R^2\int d\Omega_\mathbf{n}\,j_q^i n^i \right\rangle$.

Because the effective mass is tachyonic, the radiation contains a zero-frequency mode with $k_n=\pm m$. For point sources $\rho_*=\sum_Im_I\delta^3(\mathbf{x}-\mathbf{X}_I(t))$,
\begin{equation}
n_{(\pm 0)}(\mathbf{n}) \propto \sum_I m_I\left(\frac{1}{\mathtt{T}}\int_0^\mathtt{T}d\mathtt{t}\,e^{\mp ik_n\mathbf{n}\cdot\mathbf{X}_I}\right).
\end{equation}
It follows that $n_{(\pm 0)}(-\mathbf{n})=n_{(\pm 0)}^*(\mathbf{n})=n_{(\mp 0)}(\mathbf{n})$, so $|n_{(\pm 0)}(-\mathbf{n})|^2=|n_{(\pm 0)}(\mathbf{n})|^2$, and
\begin{equation}
\partial_k n_{(\pm 0)}(\mathbf{n}) \propto \sum_I m_I\left[\frac{\mp ik_n}{\mathtt{T}}\int_0^\mathtt{T}d\mathtt{t}\,e^{\mp ik_n\mathbf{n}\cdot\mathbf{X}_I}\frac{X_I^k-n^k(\mathbf{n}\cdot\mathbf{X}_I)}{R}\right]=-\partial_k n_{(\mp 0)}(-\mathbf{n}).
\end{equation}
The zero-frequency mode therefore does not contribute to the radiation fluxes, as it does not correspond to a propagating wave. The nonzero tachyonic mass is essential here because it shifts the zero-frequency mode to a finite wave number, avoiding the infrared divergence of the induced gravitational potential.

Having obtained the general expressions for the radiation fluxes, we now specialize to a Keplerian binary and derive more explicit results. Since the unperturbed homogeneous fluid background does not affect the binary motion, assuming a fixed Keplerian orbit amounts to requiring that the fluid perturbation induced by the binary be sufficiently weak.

\subsection{Circular orbit}\label{sec:Circular orbit}

For a Keplerian binary with $\rho_*=\sum_{I=1,2}m_I\,\delta^3(\mathbf{x}-\mathbf{X}_I(t))$, we choose a Cartesian coordinate system in which the $z$ axis aligns with the binary's orbital angular momentum and the coordinate origin is at the binary's center of mass. The fluid background is assumed to be static in this center-of-mass frame. We denote the total mass of the binary by $M\equiv m_1+m_2$ and the symmetric mass ratio by $\nu\equiv m_1m_2/M^2$.

\begin{figure}[hbt!]
\centering
\includegraphics[width=0.6\textwidth]{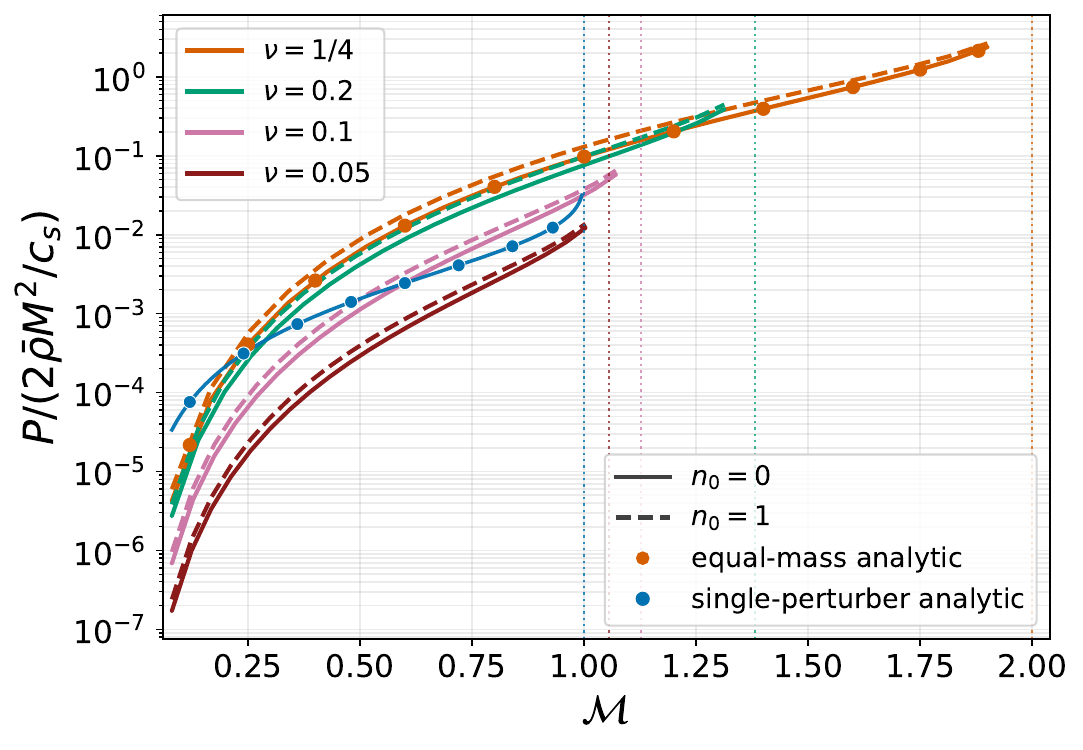}
\caption{
The dimensionless circular-orbit function $\mathcal X(\mathcal{M})$ for different symmetric mass ratios. Solid and dashed curves correspond to $n_0=0$ and $n_0=1$.  The points show the equal-mass and single-perturber analytic results, given by Eqs.~\eqref{analytic_1} and~\eqref{analytic_2}, respectively. The vertical dotted lines mark the supersonic threshold \eqref{UV}.
}\label{fig:circular_power_nu}
\end{figure}

We begin with the case of a circular orbit\footnote{In this case, it is not necessary to assume a Keplerian orbit, so the result is more broadly applicable.} and set
\begin{equation}
\begin{aligned}
\mathbf{X}_1&=R_1(\cos\Omega \mathtt{t}, \sin\Omega \mathtt{t},0),
\\
\mathbf{X}_2&=-R_2(\cos\Omega \mathtt{t}, \sin\Omega \mathtt{t},0),
\\
\mathbf{k}^{(n)}&=k_n\mathbf{n},
\\
\mathbf{n}&=(\sin\theta\cos\phi,\sin\theta\sin\phi,\cos\theta).
\end{aligned}
\end{equation}
The relative position is $\mathbf{X} \equiv\mathbf{X}_1-\mathbf{X}_2$. The two-body separation is $a=R_1+R_2$, and the velocity of the $I$th body is denoted by $\mathbf{v}_I\equiv \dot{\mathbf{X}}_I$. Using the integral representation of the Bessel function of the first kind,
\begin{equation}
J_n(z)=\int_0^{2\pi}\frac{d\xi}{2\pi}e^{i(n\xi-z\sin \xi)},
\end{equation}
and the Jacobi-Anger expansion, $e^{iy\cos z}=\sum_{n=-\infty}^\infty i^n J_n(y)e^{in z}$, we can rewrite Eq.~\eqref{source} as
\begin{equation}
\begin{aligned}
\frac{n_{(n)}}{4\pi/c_s^2} &= \sum_{I=1,2}\frac{m_I}{\mathtt{T}}\int_0^\mathtt{T}d\mathtt{t}\,e^{i\Omega_n \mathtt{t}}e^{-i\mathbf{k}^{(n)}\cdot\mathbf{X}_I}
\\
&=e^{in(\phi-\pi/2)}[m_1J_n(k_nR_1\sin\theta)+(-1)^nm_2J_n(k_nR_2\sin\theta)].
\end{aligned}
\end{equation}
From Eq.~\eqref{energy_flux}, the energy flux is given by
\begin{equation}\label{circular_P}
P=\frac{4\pi \bar\rho M^2}{c_s}\sum_{n\ge 1}\frac{\Omega_n}{k_n} I_n\equiv \frac{2\bar\rho M^2}{c_s}\mathcal{X}
,
\end{equation}
where
\begin{equation} \label{I_n}
I_n\equiv\int_0^\pi d\theta\,\sin\theta\,\left[\frac{m_1}{M}J_n(k_nR_1\sin\theta)+(-1)^n\frac{m_2}{M}J_n(k_nR_2\sin\theta)\right]^2
.
\end{equation}
We also have $P=\tau_z \tilde\Omega$, since the source has helical symmetry, $(\partial_t+\tilde\Omega\partial_\phi)\rho_*=0$. Some concrete examples are shown in Fig.~\ref{fig:circular_power_nu}. In the special case of equal masses, $m_1=m_2=M/2$, so that $R_1=R_2=a/2$, $I_n$ is nonzero only if $n$ is even, and can be obtained analytically as
\begin{equation}\label{I_n_equal}
\begin{aligned}
I_n & = \int_0^\pi d\theta\,\sin\theta\,\left[J_n\left(\frac{1}{2}k_n a\sin\theta\right)\right]^2
\\
& =\left(\frac{k_n a}{2}\right)^{2 n}\,\Gamma \left(n+\frac{1}{2}\right) \, _1\tilde{F}_2\left(n+\frac{1}{2};n+\frac{3}{2},2 n+1;-\left(\frac{k_n a}{2}\right)^2\right)
,
\end{aligned}
\end{equation}
where $_1\tilde{F}_2\left(\mathbf{a};\mathbf{b};z\right)$ is the regularized generalized hypergeometric function. We define the Mach number $\mathcal{M}\equiv a\Omega$. The velocity of each body is $v=a\tilde\Omega/2$; hence, $\mathcal{M}=2v/c_s$. For $n_0=0$, $k_n a=n\mathcal{M}$. In the subsonic regime ($\mathcal{M}<2$), the infinite series turns out to be
\begin{equation}\label{analytic_1}
\sum_{n=1}^\infty I_n(m=0)=\sum_{n=\text{even}} I_n = \frac{\operatorname{arctanh}\frac{\mathcal{M}}{2}+\ln(\tan\frac{\mathcal{E}}{2})}{\mathcal{M}}-1,
\end{equation}
where $\mathcal{E}>0$ is the solution of $\mathcal{E}-\frac{\mathcal{M}}{2}\sin \mathcal{E}=\frac{\pi}{2}$. Therefore, we obtain the tangent force
\begin{equation}
F_\text{tangent}(m=0)=\frac{P}{2v}=\frac{4\pi \bar\rho M^2}{v^2}\sum_{n\ge 1}\frac{\mathcal{M}}{4}\,I_n=\frac{4\pi \bar\rho M^2}{v^2}\left[
\frac{\operatorname{arctanh}\frac{\mathcal{M}}{2}+\ln(\tan\frac{\mathcal{E}}{2})}{4}-\frac{\mathcal{M}}{4}
\right].
\end{equation}
For $\mathcal{M}\ge 2$, the infinite series diverges. To see this, consider the large-$n$ limit, in which we may set $n_0=0$; then, for $\mathcal{M}>2$, $\lim_{n\to\infty}I_{n}\to \frac{2}{n\mathcal{M}}$. Therefore,
\begin{equation}
\lim_{n_\text{max}\to\infty}\sum_{n=\text{even}}^{n_\text{max}} I_n \sim \frac{1}{\mathcal{M}}\sum_{n=1}^{n_\text{max}} \frac{1}{n} \sim \frac{\ln n_\text{max}}{\mathcal{M}}.
\end{equation}
The energy flux thus exhibits a logarithmic UV divergence. In reality, this divergence will be regularized by the finite size of the bodies or by other short-distance physics. A simple finite-size regularization model is considered in Sec.~\ref{uniform-sphere_model}. For unequal masses, the threshold for UV divergence can be shown to be
\begin{equation}\label{UV_circular}
\frac{1+\sqrt{1-4\nu}}{2}\mathcal{M}\ge 1.
\end{equation}
Equivalently, this reads
\begin{equation}\label{UV}
\text{max}(|\mathbf{v}_1|,|\mathbf{v}_2|)=\frac{1+\sqrt{1-4\nu}}{2}|\mathbf{v}_1-\mathbf{v}_2|\ge c_s.
\end{equation}
The same supersonic threshold applies to eccentric orbits, as shown in the next subsection. An analytic expression for the energy flux appears difficult to obtain when the self-gravity of the fluid perturbation is included ($n_0>0$). In the examples shown in Fig.~\ref{fig:circular_power_nu}, self-gravity increases the energy flux.

Another analytically tractable case is the extreme-mass-ratio limit with $n_0=0$. Assuming that $m_1\gg m_2\approx \nu M$, we obtain
\begin{equation}
\frac{n_{(n)}}{4\pi/c_s^2} = \frac{\nu M}{\mathtt{T}}\int_0^\mathtt{T}d\mathtt{t}\,e^{i\Omega_n \mathtt{t}}\left[e^{i\mathbf{k}^{(n)}\cdot\mathbf{X}}-i\mathbf{k}^{(n)}\cdot\mathbf{X}\right]+\mathcal{O}(\nu^2).
\end{equation}
It follows that
\begin{equation}
P(m=0)=\frac{4\pi \bar\rho (\nu M)^2}{c_s}\left[\frac{\operatorname{arctanh}\mathcal{M}}{\mathcal{M}}-1-2\left(\frac{\sin\mathcal{M}}{\mathcal{M}}-\cos\mathcal{M}\right)+\frac{\mathcal{M}^2}{3}\right]+\mathcal{O}(\nu^3).
\end{equation}

In the single-perturber setup, a single body with mass $\nu M$ moves on a circular orbit of radius $a$ with speed $v=a\tilde\Omega$. Although this setup is unrealistic for a binary system, it isolates the dissipation associated with the gravitational wake sourced by a single body. The corresponding Mach number is $\mathcal M=a\Omega=v/c_s$. For $n_0=0$,
\begin{equation}
I_n(m=0)=\nu^2\int_0^\pi d\theta\,\sin\theta\,\left[J_n(n\mathcal M\sin\theta)\right]^2,
\end{equation}
and for $\mathcal{M}<1$, the sum gives
\begin{equation}\label{analytic_2}
\sum_{n=1}^\infty I_n(m=0) = \nu^2\left(\frac{\operatorname{arctanh}\mathcal{M}}{\mathcal{M}}-1\right),
\end{equation}
and thus
\begin{equation}
F_\text{tangent}=\frac{P}{v}=\frac{4\pi \bar\rho (\nu M)^2}{v^2}(\operatorname{arctanh}\mathcal{M}-\mathcal{M}).
\end{equation}
See Fig.~\ref{fig:circular_power_nu} for a concrete example. The dimensionless factor $\operatorname{arctanh}\mathcal M-\mathcal M=\frac12\ln[(1+\mathcal M)/(1-\mathcal M)]-\mathcal M$ coincides with Ostriker's subsonic linear-wake force factor for constant-velocity motion at a finite time after switch-on~\cite{Ostriker_1999}. The same result can be obtained by directly analyzing the force exerted on the source~\cite{Desjacques_2022,Berezhiani:2023vlo,Eytan:2025eccentric}. The absence of a large-distance logarithm in this subsonic limit should be contrasted with the logarithmic sensitivity familiar from collisionless dynamical friction and from wakes generated by a perturber moving at constant velocity in a noninteracting quantum fluid, or by a supersonic perturber in the classical fluid, where the contribution from long-wavelength modes leads to an infrared logarithmic dependence~\cite{Chandrasekhar:1943ys,Ostriker_1999,Hui:2016ltb}. In addition to the tangential gravitational force exerted by the fluid perturbation, the direct analysis reveals the force perpendicular to the velocity~\cite{Desjacques_2022}, which cannot be reconstructed from the radiation fluxes alone, since it does not enter the energy balance in the adiabatic approximation.

\subsection[Eccentric orbit]{Eccentric orbit}\label{sec:Elliptical orbit}
The radiation formulation can be extended directly to eccentric orbits with arbitrary mass ratio. For a Keplerian eccentric orbit, we can set $\mathbf{X}_1=\frac{m_2}{M}\mathbf{X}$ and $\mathbf{X}_2=-\frac{m_1}{M}\mathbf{X}$, with the parametrization:
\begin{equation}
\begin{aligned}
\mathbf{X} &=\mathbf{X}_1-\mathbf{X}_2=a\left( \cos \xi -e,\,\sqrt{1-e^2} \sin \xi ,\, 0 \right),
\\
\Omega\mathtt{t} & = \xi-e\sin \xi.
\end{aligned}
\end{equation}
If we choose $m_1\ge m_2$, then $m_1/M=(1+\sqrt{1-4\nu})/2$ and $m_2/M=(1-\sqrt{1-4\nu})/2$. The source amplitude in Eq.~\eqref{source} is evaluated as
\begin{equation}
\begin{aligned}
n_{(n)} &=\frac{4\pi}{c_s^2} \int_0^{2\pi}\frac{d\xi}{2\pi}\,(1-e\cos\xi)\,e^{in(\xi-e\sin \xi)}\left(m_1 e^{-i\frac{m_2}{M}k_n\mathbf{n}\cdot\mathbf{X}}+m_2 e^{i\frac{m_1}{M}k_n\mathbf{n}\cdot\mathbf{X}}\right)
\\
&=\frac{4\pi M}{c_s^2}K_n,
\end{aligned}
\end{equation}
with
\begin{equation}
K_n\equiv \int_0^{2\pi}\frac{d\xi}{2\pi}\,(1-e\cos\xi)\,e^{in(\xi-e\sin \xi)}\left[\frac{m_1}{M} e^{-i\frac{m_2}{M}(ak_n)\mathbf{n}\cdot(\mathbf{X}/a)}+\frac{m_2}{M} e^{i\frac{m_1}{M}(ak_n)\mathbf{n}\cdot(\mathbf{X}/a)}\right],
\end{equation}
where $n_0 \equiv m/\Omega$. From Eq.~\eqref{energy_flux}, the energy flux is given by
\begin{equation}
P= \frac{2\bar\rho M^2}{c_s}\sum_{n\ge 1} \frac{\int \sin\theta\,d\theta\,d\phi\,|K_n|^2}{\sqrt{1+(n_0/n)^2}}.
\end{equation}
From Eq.~\eqref{angular_momentum_flux}, the angular momentum flux (about the binary's center of mass) is given by
\begin{equation}
\begin{aligned}
\tau_q \tilde \Omega& =
\frac{2\bar\rho M^2}{ c_s}\sum_{|n|\ge 1}\frac{(-i)/(2n)}{\sqrt{1+(n_0/n)^2}}\int \sin\theta\,d\theta\,d\phi\,\left(\begin{matrix}
\sin\phi \,K_n\partial_\theta K_n^* + \cot \theta\, \cos \phi\, K_n\partial_\phi K_n^*
\\
-\cos\phi\,K_n\partial_\theta K_n^* + \cot \theta\, \sin \phi\, K_n\partial_\phi K_n^*
\\
-K_n\partial_\phi K^*_n
\end{matrix}\right)
\\
&=
\frac{2\bar\rho M^2}{ c_s}\text{Re}\sum_{n\ge 1}\frac{(-i)/n}{\sqrt{1+(n_0/n)^2}}\int \sin\theta\,d\theta\,d\phi\,(-K_n\partial_\phi K^*_n)\,\delta_{q,z}
.
\end{aligned}
\end{equation}
From Eq.~\eqref{linear_momentum_flux}, the linear momentum flux is given by
\begin{equation}
\begin{aligned}
\mathcal{F}_q &=
\frac{2\bar\rho M^2}{c_s^2} \sum_{n\ge 1} \int \sin\theta\,d\theta\,d\phi\,|K_n|^2n^q
\\
&=
\frac{2\bar\rho M^2}{c_s^2} \sum_{n\ge 1} \int \sin\theta\,d\theta\,d\phi\,|K_n|^2\sin\theta\,\sin\phi\,\delta_{q,y}
.
\end{aligned}
\end{equation}

\begin{figure}[hbt!]
\centering
\includegraphics[width=\textwidth]{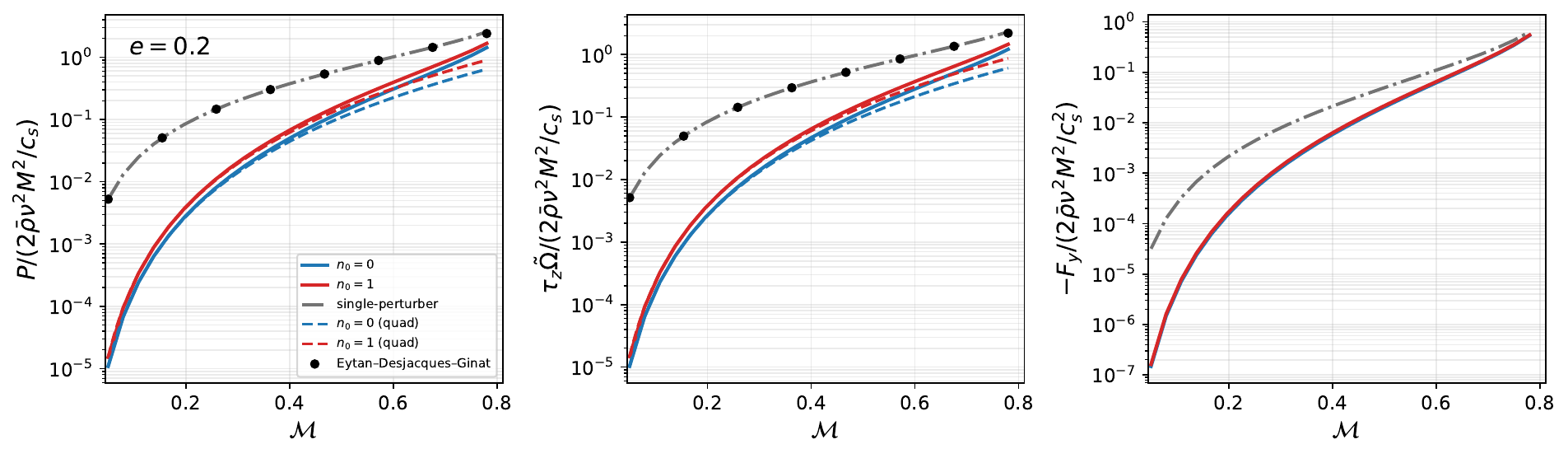}
\caption{Comparison of the full radiation fluxes with the quadrupole-order approximation~\eqref{quadrupole} in the extreme-mass-ratio limit for $e=0.2$. The panels show $P$, $\tau_z\tilde\Omega$, and $-c_s\mathcal F_y$, normalized by $2\bar\rho \nu^2M^2/c_s$. The results in the $n_0=0$ single-perturber setup are shown by the gray dot-dashed lines, which agree with the spherical-multipole series of Ref.~\cite{Eytan:2025eccentric} (marked as black points). In the single-perturber setup, a single body with mass $\nu M$ moves on an eccentric orbit with parameters $\{a,e,\tilde\Omega\}$.
}\label{fig:paper_fig1_emri_fluxes}
\end{figure}

\begin{figure}[hbt!]
\centering
\includegraphics[width=\textwidth]{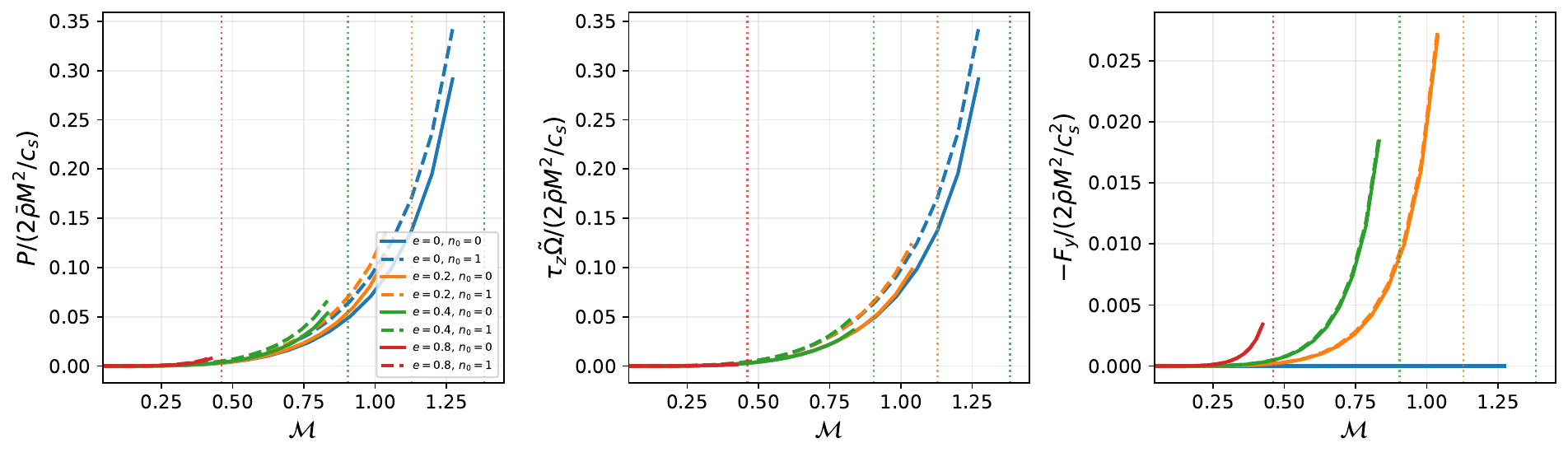}
\caption{
Sound-wave fluxes from eccentric binaries with $\nu=0.2$. The panels show $P$, $\tau_z\tilde\Omega$, and $-c_s\mathcal F_y$, normalized by $2\bar\rho M^2/c_s$. These normalized fluxes depend only on $\{e,\mathcal{M},n_0,\nu\}$.  Solid and dashed curves correspond to $n_0=0$ and $n_0=1$, respectively. The vertical dotted lines indicate the supersonic threshold \eqref{UV_elliptical}; the rapid growth near these lines signals the onset of UV divergence. The corresponding adiabatic eccentricity evolution rates are shown in the left panel of Fig.~\ref{fig:dote}.
}\label{fig:paper_fig3_nu02_ecc_fluxes}
\end{figure}

The angular-momentum flux is along the $z$ direction, parallel to the orbital angular momentum, whereas, for the cases considered here, the linear-momentum flux lies along the instantaneous velocity of the lighter component at periastron. Since the periastron of body 1 lies on the positive $x$ axis and $m_1\ge m_2$, the linear momentum flux is along the negative $y$ axis. For equal masses or a circular orbit, the linear momentum flux vanishes. See Figs.~\ref{fig:paper_fig1_emri_fluxes} and \ref{fig:paper_fig3_nu02_ecc_fluxes} for representative examples in the extreme-mass-ratio limit and for a finite-mass-ratio binary with $\nu=0.2$, respectively. In Fig.~\ref{fig:paper_fig1_emri_fluxes}, we also compare the full results with the quadrupole-order approximation (to be derived in Sec.~\ref{sec:quadrupole}), which is accurate in the deep-subsonic regime. The black points show the $n_0=0$ single-perturber results obtained by evaluating the spherical-multipole series for the time-averaged dissipated power and torque in Ref.~\cite{Eytan:2025eccentric},\footnote{Reference~\cite{Eytan:2025eccentric} also presents explicit results for eccentric binaries with arbitrary mass ratio in the $n_0=0$ case.} which were derived by directly analyzing the force exerted on the source and agree with the corresponding limit of our flux formulas.

\subsection{UV divergence from stationary phase}\label{sec:UV}

Now, we examine the UV behavior of the radiation fluxes. From Eq.~\eqref{source}, the source amplitude reads $n_{(n)}/(4\pi/c_s^2)=\sum_I m_I\mathcal{N}_I(\mathbf{n})$, with $\mathcal{N}_I(\mathbf{n})\equiv\frac{1}{T}\int_0^Tdt\,\exp\left[i n\tilde\Omega t-i k_n\mathbf n\cdot\mathbf X_I(t)\right]$. In the large-$n$ limit, $k_n/(n\tilde\Omega)\to1/c_s$.  A real stationary point exists if and only if
\begin{equation}\label{stationary_condition}
\frac{d}{dt}\left[n\tilde\Omega t-k_n\mathbf n\cdot\mathbf X_I(t)\right]=0
\quad\Longleftrightarrow\quad
\mathbf n\cdot\mathbf v_I(t)=c_s+\mathcal{O}(n^{-2}).
\end{equation}
Hence, the source has a high-frequency stationary-phase tail precisely when at least one component becomes locally supersonic somewhere on the orbit, $\max_{I,t}|\mathbf v_I(t)|> c_s$. This corresponds to Eq.~\eqref{UV} in the case of a Newtonian two-body system in the center-of-mass frame. For a Keplerian binary, $|\mathbf v_1-\mathbf v_2|_\text{max}=a\tilde\Omega \sqrt{\frac{1+e}{1-e}}$. Therefore, the exact point-source criterion is
\begin{equation}\label{UV_elliptical}
\frac{1+\sqrt{1-4\nu}}{2}
\mathcal{M}\sqrt{\frac{1+e}{1-e}}> 1,
\end{equation}
with $\mathcal{M}=a\Omega$. In the strictly subsonic regime, the phase has no real stationary point, and the Fourier harmonics at large-$n$ are rapidly suppressed.  When Eq.~\eqref{UV_elliptical} is satisfied, the stationary phase gives $K_n\sim n^{-1/2}$ over an open set of emission directions.  Consequently, $\int d\Omega_\mathbf{n} |K_n|^2\sim n^{-1}$, and the radiation fluxes of point particles acquire a logarithmic dependence on the UV cutoff,\footnote{Note that, for $0<e<1$, the fluxes remain UV convergent when $\max_{I,t}|\mathbf v_I(t)|=c_s$.} indicating the failure of the theory at sufficiently short distances. We note that, since the time-averaged radiation fluxes are quadratic in the source amplitude $n_{(n)}\propto\sum_I m_I\mathcal{N}_I$, the interference terms in the fluxes arise from the so-called ``cross-wakes,'' i.e., the influence of the wake sourced by one body on the other bodies. The analysis above shows that the UV divergence is associated with the self-wake, while the cross-wake contributions remain UV convergent.

\subsection{Simple model of UV regularization}\label{uniform-sphere_model}
The UV divergence in the supersonic case is a pathology of the point-particle approximation within linear response theory. It is removed, for example, when the perturbers have finite size. In this section, we consider a simple model of this type. Still assuming that linear perturbation theory applies and that the centers of the bodies follow prescribed periodic orbits $\mathbf{X}_I(t)$, but dropping the point-particle approximation, we write
\begin{equation}
\rho_*(t,\mathbf{x})=\sum_I m_I \mathcal{W}_I(\mathbf{x}-\mathbf{X}_I(t)),
\end{equation}
and consequently,
\begin{equation}
\frac{n_{(n)}}{4\pi/c_s^2} = \sum_{I=1,2}\frac{m_I}{\mathtt{T}}\int_0^\mathtt{T}d\mathtt{t}\,e^{i\Omega_n \mathtt{t}}e^{-i\mathbf{k}^{(n)}\cdot\mathbf{X}_I}W_I(\mathbf{k}^{(n)}),
\end{equation}
with the form factor
\begin{equation}
W_I(\mathbf{k})= \int d^3x\,\mathcal{W}_I(\mathbf{x})\,e^{-i\mathbf{k}\cdot\mathbf{x}}.
\end{equation}
The point-particle approximation corresponds to $W_I=1$.

Here, we consider a uniform-sphere model, defined by $\mathcal{W}_I(\mathbf{x})=\frac{3}{4\pi R_I^3}\Theta(R_I-|\mathbf{x}|)$, where $\Theta(z)$ is the Heaviside unit step function. This corresponds to treating each body as a sphere of uniform mass density and radius $R_I$. The corresponding form factor is given by
\begin{equation}
W_I(\mathbf{k})=3\frac{\sin(kR_I)-kR_I\cos(kR_I)}{(kR_I)^3}, \label{W}
\end{equation}
where $k=|\mathbf{k}|$. As $k\to 0$, $W_I(k)=1-(kR_I)^2/10+(kR_I)^4/280+\mathcal{O}\left((kR_I)^6\right)$. For a Newtonian binary system, a desirable aspect of this model is that, provided the two bodies remain spherically symmetric and nonoverlapping, the unperturbed orbits of the two centers remain exactly Keplerian by Newton's shell theorem. Because of the short-wavelength suppression of $W_I(k)$, the radiation fluxes in this model are UV finite. For $R_I=R$, the effective UV cutoff introduced by the form factor \eqref{W} in the point-particle theory can be shown to be
\begin{equation}
n^\text{eff}_\text{max}=\left\lfloor \frac{e^{7/4-\gamma_E}}{2(R/a)\mathcal{M}} \right\rfloor,\label{n^eff}
\end{equation}
where $\lfloor x\rfloor$ denotes the floor function and $\gamma_E$ is the Euler-Mascheroni constant.

\begin{figure}[hbt!]
\centering
\includegraphics[width=0.6\textwidth]{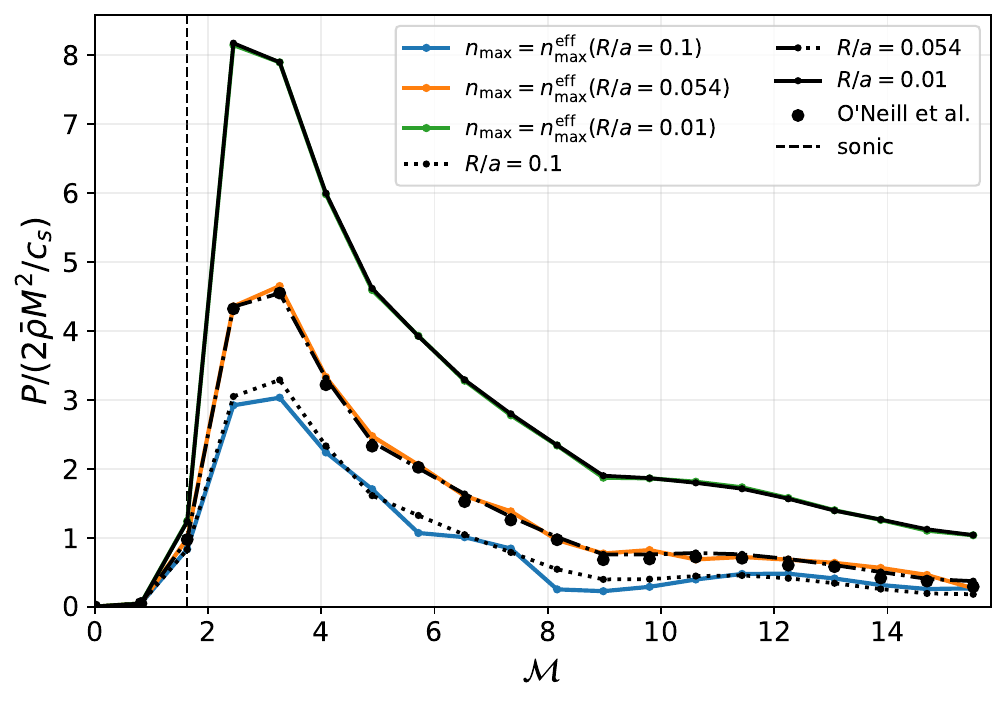}
\caption{Dimensionless sound-wave energy flux for an equal-mass binary ($\nu=1/4$) with $e=0.2$ and $n_0=0$. Large black points show the numerical results from the direct time-domain wake calculation of O'Neill et al.~\cite{2024ApJ...974..216O}. The black dotted, dot-dashed, and solid lines show the uniform-sphere results for $R_1=R_2=R$, with $R/a=0.1$, $0.054$, and $0.01$, respectively. Colored lines show the point-source harmonic sums truncated at $n_{\max}=n^\text{eff}_\text{max}(R)$. Note that at $\mathcal{M}=14$, $n^\text{eff}_\text{max}(R/a=0.1)=1$, $n^\text{eff}_\text{max}(R/a=0.054)=2$, and $n^\text{eff}_\text{max}(R/a=0.01)=11$. The vertical black dashed line indicates the supersonic threshold \eqref{UV_elliptical}.
}\label{fig:oneill_binary_e02_power_comparison}
\end{figure}

In Fig.~\ref{fig:oneill_binary_e02_power_comparison}, we compare the energy fluxes obtained in this model with the point-particle results evaluated using the effective hard cutoff $n\le n_{\max}^{\rm eff}$ for a concrete example. The effective hard-cutoff prescription is found to provide a good approximation to the uniform-sphere results for sufficiently small sphere radii. Remarkably, the results for $R/a=0.054$ agree well with the direct time-domain wake calculation of O'Neill et al.~\cite{2024ApJ...974..216O}, which implements a different short-distance regularization by excising a small region around each perturber. We note that Ref.~\cite{Eytan:2025eccentric} employs another numerical UV regularization in the supersonic regime by truncating the spherical-harmonic expansion of the point-particle results.

\subsection{Quadrupole-order approximation as subsonic limit}
\label{sec:quadrupole}
We now return to a UV-finite regime of the point-particle theory. If $ak_n \ll 1$, we can use a long-wavelength expansion of $e^{-i\mathbf{k}^{(n)}\cdot\mathbf{X}_I}$ in Eq.~\eqref{source}, and the leading term corresponds to the quadrupole-order radiation. For radiation with a relativistic wave speed, the long-wavelength approximation is justified for nonrelativistic binaries because $ak_n$ is of order the orbital velocity. In the present case, however, $a k_n=\mathcal{M}\sqrt{n^2+n_0^2}$. Thus, the long-wavelength expansion requires $\mathcal{M}\sqrt{n^2+n_0^2}\ll1$ for the harmonics under consideration. For sufficiently small $n_0$, this reduces to the deep-subsonic condition $\mathcal{M}\ll1$. In this regime, we can use the quadrupole-order approximation,
\begin{equation}
\begin{aligned}
\frac{n_{(n)}}{4\pi/c_s^2} &= \sum_{I=1,2}\frac{m_I}{\mathtt{T}}\int_0^\mathtt{T}d\mathtt{t}\,e^{i\Omega_n \mathtt{t}}e^{-i\mathbf{k}^{(n)}\cdot\mathbf{X}_I}
\\
&\approx \frac{n_{(n)}^{[2]}}{4\pi/c_s^2}\equiv
-\sum_I\frac{m_I k_n^2}{2}\frac{1}{\mathtt{T}}\int_0^\mathtt{T}d\mathtt{t}\,e^{i\Omega_n \mathtt{t}} n^i n^j X_I^i X_I^j
\\
&=
-\frac{\nu M k_n^2}{2}\mathcal{I}_{ij}n^i n^j
,
\end{aligned}
\end{equation}
with
\begin{equation}
\mathcal{I}_{ij}=\frac{1}{\mathtt{T}}\int_0^\mathtt{T}d\mathtt{t}\,e^{i\Omega_n \mathtt{t}} X^i X^j
=
\int_0^{2\pi}\frac{d\xi}{2\pi}\,(1-e\cos\xi)\,e^{in(\xi-e\sin \xi)} X^i X^j.
\end{equation}
Here, the mass-dipole term vanishes in the center-of-mass frame, while the leading $\mathcal{O}(k^2)$ contribution is determined by the second mass moment $I_{ij}=\sum_I m_I X_I^iX_I^j$, whose trace and symmetric trace-free parts correspond to the $l=0$ and $l=2$ sectors of the spherical-harmonic decomposition, respectively, i.e., to monopolar and quadrupolar radiation; for a circular orbit, the trace is time independent, so only the latter radiates. Note that for $n\ge 1$
{
\small
\begin{equation}
\begin{aligned}
&|\mathcal{I}_{kk}|^2+2 \mathcal{I}_{ij}\mathcal{I}_{ij}^*
\\
&=
\frac{4 a^4 \left\{4 e^2 \left(e^2-1\right) \left[\left(e^2-1\right) n^2-1\right] (J'_n)^2-4 e \left(3 e^4-7 e^2+4\right) n J_n J_n'+\left[3 e^4-4 \left(e^2-1\right)^3 n^2-4 e^2+4\right] (J_n)^2\right\}}{e^4 n^4}
,
\end{aligned}
\end{equation}
\begin{equation}
\epsilon_{qkl} \mathcal{I}_{ki}^* \mathcal{I}_{li}=\delta_{q,z}
\frac{8 i a^4 \sqrt{1-e^2} \left[\left(e^2-1\right) n J_n+e J_n'\right] \left[2 e \left(e^2-1\right) n J'_n-\left(e^2-2\right) J_n\right]}{e^4 n^4}
,
\end{equation}
}\noindent where $J_n\equiv J_n(ne)$, $J_n'\equiv \left[{dJ_n(z)}/{dz}\right]_{z=ne}=(J_{n-1}-J_{n+1})/2$. Define
\begin{equation}
n_{(n)}^{[2]}=-\frac{4\pi}{c_s^2}\frac{\nu M k_n^2}{2}\,Q_n(\mathbf n),\qquad Q_n(\mathbf n)\equiv \mathcal I_{ij}n^in^j;
\end{equation}
the required angular integrals are then
\begin{equation}
\begin{aligned}
\int d\Omega_\mathbf n\,|Q_n|^2
&=\mathcal I_{ij}\mathcal I_{kl}^*\int d\Omega_\mathbf n\, n^i n^j n^k n^l
=\frac{4\pi}{15}\left(|\mathcal I_{kk}|^2+2\mathcal I_{ij}\mathcal I_{ij}^*\right),
\\
\int d\Omega_\mathbf n\,\epsilon_{qkl}x^l Q_n\partial_kQ_n^*
&=R\mathcal I_{ij}\mathcal I_{sp}^*\int d\Omega_\mathbf n\,\epsilon_{qkl}n^l n^i n^j\partial_k(n^s n^p)
=\frac{16\pi}{15}\epsilon_{qkl}\mathcal I_{ki}^*\mathcal I_{li}.
\end{aligned}
\end{equation}
Restoring the prefactor in $n_{(n)}^{[2]}$ in Eqs.~\eqref{energy_flux} and~\eqref{angular_momentum_flux} yields $\boldsymbol{\mathcal{F}=0}$, and
\begin{equation}\label{quadrupole}
\begin{aligned}
P & = \sum_{n\ge 1} \frac{2\pi}{15} \frac{\nu^2M^2 \bar\rho k_n^3\Omega_n}{c_s}\left(	|\mathcal{I}_{kk}|^2+2 \mathcal{I}_{ij}\mathcal{I}_{ij}^*\right)
,
\\
\boldsymbol{\tau} &= \tau_z\,\mathbf{e}_z =\sum_{n\ge 1} \frac{8\pi}{15} \frac{\nu^2M^2 \bar\rho k_n^3}{c_s^2}\left(-i
\epsilon_{zkl} \mathcal{I}^*_{ki} \mathcal{I}_{li}
\right)\,\mathbf{e}_z
.
\end{aligned}
\end{equation}
For a circular orbit ($e=0$),
\begin{align}
|\mathcal{I}_{kk}|^2+2 \mathcal{I}_{ij}\mathcal{I}_{ij}^* & = \frac{a^4}{2}(\delta_{n,2}+\delta_{n,-2})
,
\quad
\epsilon_{zkl} \mathcal{I}^*_{ki} \mathcal{I}_{li}  =  i\frac{a^4}{4} (\delta_{n,2}-\delta_{n,-2})
;
\end{align}
hence
\begin{equation}
\begin{aligned}
P = \frac{2\pi}{15} \frac{\nu^2M^2 \bar\rho k_2^3\Omega a^4}{c_s} =\tau_z \tilde\Omega
.
\end{aligned}
\end{equation}
This agrees with the long-wavelength limit of Eq.~\eqref{circular_P}.

If the fluid self-gravity is neglected by setting $n_0=0$, we obtain
\begin{equation}
\begin{aligned} \label{m=0_flux}
P & = \frac{4\pi}{15} \frac{\nu^2M^2 \bar\rho \Omega^4a^4}{c_s}\frac{7 \sqrt{1-e^2}-3(1-e^2)}{1-e^2}
,
\\
\tau_z &= \frac{16\pi}{15} \frac{\nu^2M^2 \bar\rho \Omega^3a^4 }{c_s^2}\sqrt{1-e^2}
.
\end{aligned}
\end{equation}
If the binary evolves adiabatically under the associated radiation reaction, with $\tilde\Omega^2=\tilde M/a^3$ and constant $\tilde M$ (not necessarily equal to $M$; henceforth we set $\tilde M=M$ unless stated otherwise), the secular rates of change of the osculating Keplerian orbital elements $\{a,e\}$ are given by
\begin{equation}\label{orbital_evolution}
\begin{aligned}
\dot a &=-\frac{2a^2P}{\nu M\tilde M},
\\
\dot e &=-\left(\frac{1-e^2}{e}\right)\frac{a}{\nu M\tilde M}\left(P-\frac{\tau_z \tilde\Omega}{\sqrt{1-e^2}}\right).
\end{aligned}
\end{equation}
For Eq.~\eqref{m=0_flux}, we find that
\begin{equation}
P-\frac{\tau_z \tilde\Omega}{\sqrt{1-e^2}}
=
\frac{28\pi}{15} \frac{\nu^2M^2 \bar\rho \Omega^4a^4}{c_s}
\left(\frac{1}{\sqrt{1-e^2}}-1\right)>0
.
\end{equation}
The sound radiation in this case therefore tends to circularize the orbit.

\begin{figure}[hbt!]
\centering
\includegraphics[width=\textwidth]{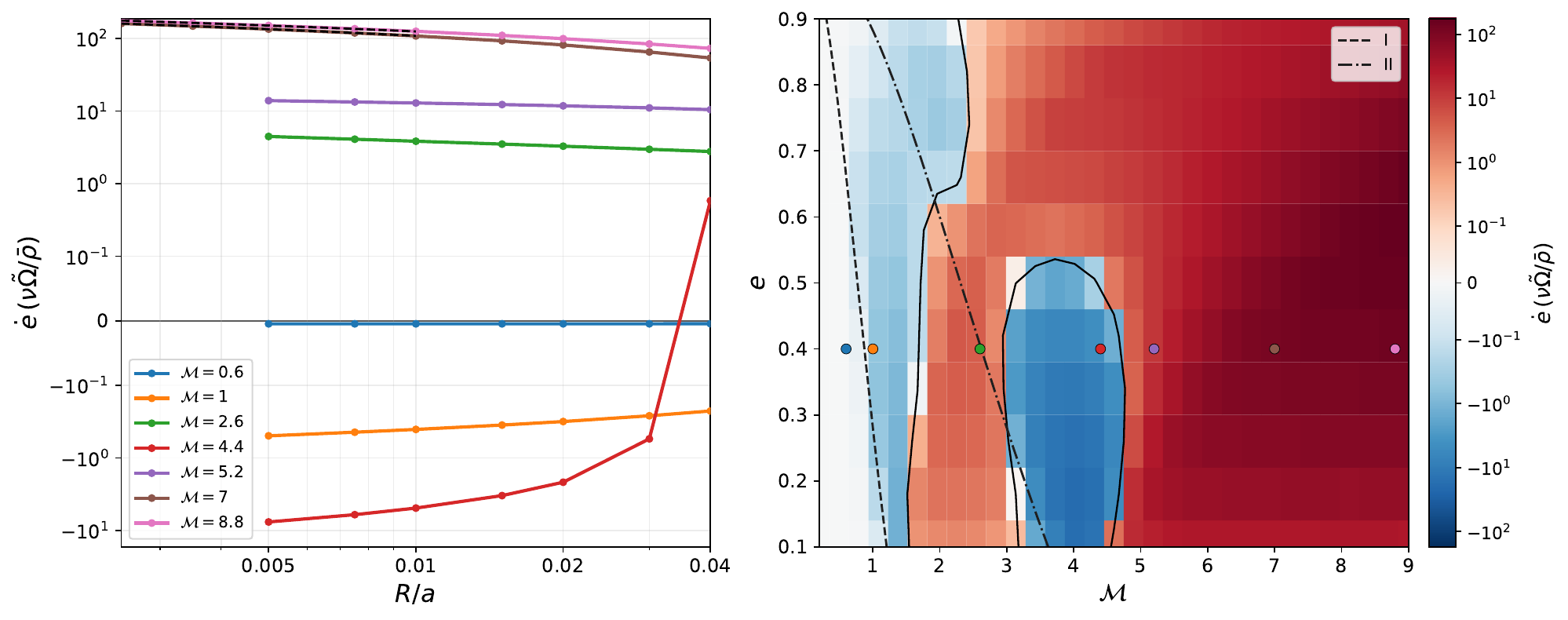}
\caption{Adiabatic eccentricity evolution rate $\dot e$ (normalized by $\frac{\bar\rho}{\nu\tilde\Omega}$) for eccentric binaries with $\nu=3/16$ and $n_0=0$. Left: results for $e=0.4$ in the uniform-sphere model. The black dashed lines show the analytic small-$R$ logarithmic asymptotes of Eq.~\eqref{dote_small_R}, with their additive constants determined by matching at $R=0.005a$. Right: $\dot e$ in the $(e,\mathcal M)$ plane for $R=0.01a$, with the colored markers corresponding to the cases shown in the left panel. The solid black contour marks $\dot e=0$. The lines labeled I and II mark the thresholds at which one body and both bodies, respectively, become supersonic at periapsis, corresponding to $\frac{1+\sqrt{1-4\nu}}{2}\mathcal{M}\sqrt{\frac{1+e}{1-e}}=1$ and $\frac{1-\sqrt{1-4\nu}}{2}\mathcal{M}\sqrt{\frac{1+e}{1-e}}=1$.
}\label{fig:classical_eccentricity_growth}
\end{figure}

For sufficiently small $n_0$, our numerical results indicate that $\dot e<0$ throughout the strictly subsonic regime, although a general analytic proof appears to be difficult. As was observed in Refs.~\cite{2024ApJ...974..216O,Eytan:2025eccentric}, outside the subsonic regime $\dot e$ may become positive. To study this regime, we adopt the uniform-sphere model presented in Sec.~\ref{uniform-sphere_model}, with $R_1=R_2=R$. The resulting behavior is illustrated in Fig.~\ref{fig:classical_eccentricity_growth}, which shows both the $R$ dependence of $\dot e$ and its value at fixed $R$ across the $(e,\mathcal M)$ plane for a concrete example. In the strictly supersonic regime of the point-particle theory, where both bodies remain supersonic throughout the orbit, we find that $\dot e(n_{\max})=C(e,\nu)\ln n_{\max}+\mathcal{O}(1)$ as $n_{\max}\to\infty$, where
\begin{equation}
C(e,\nu) = \left[-\frac{4(1-e^2)(1-3\nu)}{\nu^2} \int_0^{2\pi}d\xi\, \cos\xi \left( \frac{1-e\cos\xi}{1+e\cos\xi} \right)^{3/2}\right]\frac{\bar\rho}{\tilde\Omega}>0. \label{C_factor}
\end{equation}
The leading logarithmic coefficient is independent of $n_0$ for $n_{\max}\gg n_0$. Hence, for sufficiently large $n_{\max}$, strictly supersonic motion necessarily drives eccentricity growth for arbitrary mass ratio. For sufficiently small $n_0$, combining this result with the negative $\dot e$ found throughout the strictly subsonic regime implies a crossover in the sign of $\dot e$ at an intermediate Mach number. Combined with Eq.~\eqref{n^eff}, the large-$n_{\max}$ result further implies that, in the strictly supersonic regime of the uniform-sphere model,
\begin{equation}
\dot e(R)=-C(e,\nu)\ln\left[(R/a)\mathcal M\right]+\mathcal O(1) \label{dote_small_R}
\end{equation}
as $R/a\to0$ at fixed $\mathcal M$ and $n_0$.

\begin{table}[!h]
\centering
\footnotesize
\renewcommand{\arraystretch}{1.25}
\begin{tabular}{c@{\hspace{2cm}}c}
\toprule
\textbf{Classical fluid} &
\textbf{Quantum fluid} \\
\midrule
$\mathtt{t}=c_s t$ &
$\mathtt{t}=t/(2m_\phi)$ \\
$\Omega=\tilde\Omega/c_s$ &
$\Omega=2m_\phi\tilde\Omega$ \\
$m=\sqrt{4\pi\bar\rho}/c_s$ &
$m=2m_\phi\sqrt{4\pi\bar\rho}$ \\
$n(\mathtt{t},\mathbf{x})=\frac{4\pi}{c_s^2}\rho_*$ &
$n(\mathtt{t},\mathbf{x})=16\pi m_\phi^2 \,\rho_*$
\\
\multicolumn{2}{c}{$n_0=m/\Omega=\sqrt{4\pi\bar\rho}/\tilde\Omega$} \\
$\mathcal{M}=a\Omega$ &
$\mathcal{M}_Q=a\sqrt{\Omega}$ \\
\bottomrule
\end{tabular}
\caption{Notation used in the classical- and quantum-fluid formulations. Here, $\tilde\Omega=2\pi/T$, $c_s$, and $m_\phi$ denote the orbital angular frequency of the Keplerian binary, classical-fluid sound speed, and scalar boson mass, respectively.}
\label{tab:notation}
\end{table}

\section{Binary in a quantum fluid}\label{sec:quantum}
The nonrelativistic limit of a massive fundamental field also admits an effective description as a self-gravitating Newtonian fluid, which is relevant to ultralight scalar dark matter models~\cite{Hui:2016ltb}. Consider a real, canonically normalized scalar field $\phi$ with the potential,
\begin{equation}
V(\phi)=\frac{1}{2}m_\phi^2\phi^2 + \frac{1}{4}\lambda \phi^4,
\end{equation}
where $m_\phi$ is the scalar mass and $\lambda$ is the quartic self-coupling. In the nonrelativistic limit, the slow mode of the scalar field in a flat spacetime satisfies the Schr{\"o}dinger-Poisson (SP) equation~\cite{Salehian_2021},
\begin{equation}
\begin{aligned}
\nabla^2\varPhi &= 4\pi \left(m_\phi|\psi|^2-\bar\rho+\rho_*\right),
\\
i\dot\psi &= -\frac{1}{2m_\phi}\nabla^2\psi + m_\phi\varPhi \psi+\frac{3\lambda}{4m_\phi^2}|\psi|^2\psi,
\end{aligned}
\end{equation}
with
\begin{equation}
\phi=\frac{1}{\sqrt{2m_\phi}}(\psi\,e^{-i m_\phi t}+\text{c.c.}),
\end{equation}
where $\psi$ is the complex scalar-field amplitude associated with the slowly varying mode. As in Sec.~\ref{sec:1}, the Jeans prescription is adopted, and $\rho_*$ denotes the mass density external to the fluid. The Madelung transformation recasts the wave function as a quantum fluid.\footnote{Despite the name, this is a classical mean-field description of the scalar condensate. A genuinely quantum many-body description in the nonrelativistic regime would promote $\psi$ to a field operator. For a highly occupied condensate (as in the astrophysical scenarios of interest), a coherent state provides an appropriate description, whose expectation value defines the classical field. The interesting problem of nonclassical effects is beyond the scope of this paper.}  We introduce the variables $\{\rho,\mathbf{v}\}$ by $\psi\equiv\sqrt{\frac{\rho}{m_\phi}}\,e^{im_\phi s}$ and $\mathbf{v}\equiv\nabla s=\frac{i}{2\rho}(\psi\nabla\psi^*-\psi^*\nabla\psi)$. The SP equations become
\begin{align}
\dot\rho + \nabla\cdot(\rho\mathbf{v})&=0,
\\
\dot{\mathbf{v}}+\mathbf{v}\cdot\nabla\mathbf{v}
&=\frac{1}{2m_\phi^2}\nabla\left(\frac{\nabla^2\sqrt{\rho}}{\sqrt{\rho}}\right)-\nabla\left(\varPhi+\frac{\rho}{\rhoint}\right), \label{quantum_Euler}
\\
\nabla^2\varPhi &=4\pi(\rho-\bar\rho+\rho_*)
,
\end{align}
where $\rhoint\equiv 4m_\phi^4/(3\lambda)$, and the first term on the r.h.s. of Eq.~\eqref{quantum_Euler}, $\frac{1}{2m_\phi^2}\partial_i\left(\frac{\nabla^2\sqrt{\rho}}{\sqrt{\rho}}\right)=-\frac{1}{\rho}\partial_j\left(-\frac{\rho \partial_j\partial_i\ln \rho}{4m_\phi^2}\right)$, is the contribution from the so-called quantum pressure. For $\rho_*=0$, a static homogeneous background solution is given by $\nabla\bar\rho=\nabla\bar\varPhi=\bar{\mathbf{v}}=\mathbf{0}$. As in Sec.~\ref{sec:1}, we are interested in the linear perturbation of this fluid background by a set of nonrelativistic point masses, corresponding to the perturber density $\rho_*$. We decompose the fields as
\begin{equation}
\rho=\bar\rho(1+\alpha),
\quad
\varPhi=\bar\varPhi + \Phi, \quad
\mathbf{v}=\bar{\mathbf{v}}+\mathbf{u}.
\end{equation}
The linearized fluid equations then read
\begin{align}
\nabla^2\Phi &=
4\pi(\bar\rho\,\alpha+\rho_*),
\\
\dot{\mathbf{u}} &= -\nabla\Phi-\frac{\bar\rho }{\rhoint}\nabla\alpha+\frac{1}{4m_\phi^2}\nabla^3\alpha
,
\\
\dot\alpha+\nabla\cdot\mathbf{u} &=0.
\end{align}
It follows that
\begin{equation}
\ddot\alpha-\left(\frac{\bar\rho }{\rhoint}\right)\nabla^2\alpha+\frac{1}{4m_\phi^2}\nabla^4\alpha=4\pi (\bar\rho\, \alpha+\rho_*). \label{quantum_sound}
\end{equation}
We note that the same result holds in the wave perturbation theory. Writing $\psi=\sqrt{\frac{\bar\rho}{m_\phi}} e^{-i\mu t }\left(1+u+iv\right)$, with the chemical potential $\mu=\frac{m_\phi}{\rhoint}\bar\rho$, we obtain the Bogoliubov--de Gennes equations
\begin{equation}
\dot u=-\frac{1}{2m_\phi}\nabla^2 v,
\quad
\dot v=\frac{1}{2m_\phi}\nabla^2 u-\frac{2m_\phi\bar\rho}{\rhoint}u-m_\phi\Phi,
\end{equation}
which gives Eq.~\eqref{quantum_sound}, with the identification $u=\alpha/2$. The phase perturbation $v$, on the other hand, is driven by a nonlocal source term.

\subsection{Noether currents}
As in the classical-fluid case, we now turn to a Lagrangian formulation. The Lagrangian for $\{s,\alpha,\Phi\}$ can be taken as
\begin{equation}
\mathcal{L}=-\frac{\bar\rho }{2}|\nabla s|^2-\frac{\bar\rho ^2}{2\rhoint}\alpha^2 - \bar\rho \Phi\alpha - \frac{\bar\rho }{8m_\phi^2}|\nabla\alpha|^2 + \bar\rho s\dot \alpha-\frac{1}{8\pi}|\nabla\Phi|^2-\rho_*\Phi.
\end{equation}
From Noether's theorem (setting $\rho_*=0$), we obtain the energy flux density
\begin{equation}\label{quantum_energy_flux}
\mathbf{S}=\frac{\partial\mathcal{L}}{\partial \nabla s}\dot s+\frac{\partial\mathcal{L}}{\partial \nabla \alpha}\dot \alpha+\frac{\partial\mathcal{L}}{\partial \nabla \Phi}\dot \Phi=-\bar\rho  \dot s\nabla s-\frac{\bar\rho}{4m_\phi^2}  \dot \alpha\nabla \alpha-\frac{1}{4\pi}\dot\Phi\nabla\Phi
,
\end{equation}
the energy density
\begin{equation}
\mathcal{E} = \frac{\partial\mathcal{L}}{\partial \dot \alpha}\dot \alpha-\mathcal{L}=\bar\rho s\dot\alpha-\mathcal{L},
\end{equation}
the linear momentum flux density
\begin{equation}
\begin{aligned}
\Pi_q^i &=-\left[\frac{\partial \mathcal{L}}{\partial (\partial_i s)}\partial_qs
+\frac{\partial \mathcal{L}}{\partial (\partial_i\Phi)}\partial_q\Phi
+\frac{\partial \mathcal{L}}{\partial (\partial_i\alpha)}\partial_q\alpha
-\delta_{iq}\mathcal{L}\right]
\\
&=\bar\rho \partial_i s \partial_q s
+\frac{1}{4\pi}\partial_i\Phi \partial_q\Phi
+\frac{\bar \rho}{4m_\phi^2}\partial_i\alpha \partial_q\alpha
+\delta_{iq} \mathcal{L}
,
\end{aligned}
\end{equation}
the linear momentum density
\begin{equation}
\begin{aligned}
p_q =-\frac{\partial \mathcal{L}}{\partial \dot\alpha}\partial_q\alpha
=-\bar\rho s \partial_q \alpha
,
\end{aligned}
\end{equation}
the angular momentum flux density
\begin{equation}
\begin{aligned}
j_q^i &=\frac{1}{2}\epsilon_{qkl}\left[
\frac{\partial \mathcal{L}}{\partial (\partial_i s)}A^j _{kl}\partial_j s
+\frac{\partial \mathcal{L}}{\partial (\partial_i \alpha)}A^j _{kl}\partial_j \alpha
+\frac{\partial \mathcal{L}}{\partial (\partial_i \Phi)}A^j _{kl}\partial_j \Phi -A^i _{kl}\mathcal{L}
\right]
\\
&=\frac{1}{2}\epsilon_{qkl}\left[
-\bar\rho \partial_i s \left(x^l\partial_k-x^k\partial_l\right)s-\frac{\bar\rho}{4m_\phi^2}\partial_i\alpha\left(x^l\partial_k-x^k\partial_l\right)\alpha\right.\\
&\qquad\qquad \left.-\frac{1}{4\pi}\partial_i\Phi\left(x^l\partial_k-x^k\partial_l\right)\Phi
-(\delta_{ik}x^l-\delta_{il}x^k)\mathcal{L}
\right]
,
\end{aligned}
\end{equation}
and the angular momentum density
\begin{equation}
\begin{aligned}
j_q = \frac{1}{2}\epsilon_{qkl}\left[
\frac{\partial \mathcal{L}}{\partial \dot \alpha}A^j _{kl}\partial_j \alpha
\right]
= \bar\rho\epsilon_{qkl} x^l s\partial_k \alpha.
\end{aligned}
\end{equation}

Equation~\eqref{quantum_energy_flux} also follows directly from the underlying wave function theory.  With the point-particle mass density $\rho_*(t,\mathbf x)=\sum_I m_I\delta^3(\mathbf x-\mathbf X_I(t))$, a convenient Newtonian Lagrangian density is
\begin{equation}
\mathcal{L}_\psi=\frac{i}{2}\left(\psi^*\dot\psi-\dot\psi^*\psi\right)-\frac{1}{2m_\phi}|\nabla\psi|^2-\frac{3\lambda}{8m_\phi^2}|\psi|^4-m_\phi\Phi|\psi|^2-(\rho_*-\bar\rho)\Phi-\frac{1}{8\pi}|\nabla\Phi|^2+\sum_I\frac{m_I}{2}|\dot{\mathbf X}_I|^2\delta^3(\mathbf x-\mathbf X_I).
\end{equation}
With $\psi=\sqrt{\rho/m_\phi}\,e^{im_\phi s}$, the spatial-gradient terms give the energy flux
\begin{equation}
\begin{aligned}
\mathbf S&=\frac{\partial\mathcal L_\psi}{\partial\nabla\psi}\dot\psi+\frac{\partial\mathcal L_\psi}{\partial\nabla\psi^*}\dot\psi^*+\frac{\partial\mathcal L_\psi}{\partial\nabla\Phi}\dot\Phi
\\
&=-\frac{1}{2m_\phi}(\dot\psi\nabla\psi^*+\dot\psi^*\nabla\psi)-\frac{1}{4\pi}\dot\Phi\nabla\Phi
\\
&=-\rho\dot s\nabla s-\frac{1}{4m_\phi^2}\frac{\dot\rho\nabla\rho}{\rho}-\frac{1}{4\pi}\dot\Phi\nabla\Phi.
\end{aligned}
\end{equation}
Expanding to quadratic order in the perturbations reproduces Eq.~\eqref{quantum_energy_flux}.

\subsection{Radiation from a localized periodic source}
Introducing a new time coordinate $\mathtt{t}=t/(2m_\phi)$, we rewrite Eq.~\eqref{quantum_sound} as
\begin{equation}\label{redefined_sound_2}
\partial_{\mathtt t}^2\alpha-c_S^2\nabla^2\alpha+\nabla^4\alpha - m^2\alpha =n(\mathtt{t},\mathbf{x})\equiv 16\pi m_\phi^2\,\rho_*(t=2m_\phi\mathtt t,\mathbf{x}),
\end{equation}
with
\begin{equation}
m\equiv \sqrt{16\pi m_\phi^2\bar\rho},
\quad
c_S^2\equiv\frac{4m_\phi^2\bar\rho }{\rhoint}.
\end{equation}
The dispersion relation of a free plane wave $\alpha\propto e^{i(\mathbf{k\cdot x}-\omega \mathtt{t})}$ is $\omega^2=c_S^2|\mathbf{k}|^2+|\mathbf{k}|^4-m^2$. For a repulsive quartic self-interaction ($\rhoint=4m_\phi^4/(3\lambda)>0$) and negligible quantum pressure (i.e., neglecting the $\nabla^4$ term in the wave equation\footnote{This approximation is justified for radiation modes satisfying $k\ll c_S$ (corresponding to wavelengths much longer than the healing length $\xi_h=\sqrt{2}/c_S$), so that the quantum-pressure term is negligible.}), the quartic self-interaction term gives rise to an effective barotropic pressure analogous to that in the classical fluid considered in Sec.~\ref{sec:1}; see Eq.~\eqref{redefined_sound}. For an attractive or vanishing quartic self-interaction, the quantum pressure is always important for radiation.

In the frequency domain, Eq.~\eqref{redefined_sound_2} reads $(\nabla^4-c_S^2\nabla^2-m^2-\omega^2)\tilde\alpha=\tilde n$. For positive frequency $\omega>0$, the outgoing solution to $(\nabla^4-c_S^2\nabla^2 -m^2 - \omega^2) f(\mathbf{x},\mathbf{x}')=\delta^3(\mathbf{x}-\mathbf{x}')$ is
\begin{equation}
\begin{aligned}
f(\mathbf{x},\mathbf{x}')&=\int\frac{d^3k}{(2\pi)^3} \frac{e^{i\mathbf{k\cdot (x-x')}}}{|\mathbf{k}|^4+c_S^2|\mathbf{k}|^2-m^2-(\omega+i\,0^+)^2}
\\
&=\frac{e^{ iq_+|\mathbf{x}-\mathbf{x}'|}-e^{-q_-|\mathbf{x}-\mathbf{x}'|}}{4\pi(q_+^2+q_-^2)|\mathbf{x}-\mathbf{x}'|}
\\
&\overset{|\mathbf{x}-\mathbf{x}'|\to\infty}{\to} \frac{1}{\sqrt{c_S^4+4(\omega^2+m^2)}}\frac{e^{iq_+|\mathbf{x}-\mathbf{x}'|}}{4\pi|\mathbf{x}-\mathbf{x}'|},
\end{aligned}
\end{equation}
where
\begin{equation}
q_\pm = \sqrt{ \frac{\mp c_S^2+\sqrt{c_S^4+4(\omega^2+m^2)}}{2} }.
\end{equation}
In the limit $(\omega^2+m^2)/c_S^4\to 0$, $q_+\to\sqrt{\omega^2+m^2}\,/c_S$, $f(\mathbf{x},\mathbf{x}')\to \frac{1}{c_S^2}\frac{e^{i\left(\sqrt{\omega^2+m^2}\,/c_S\right)|\mathbf{x}-\mathbf{x}'|}}{4\pi|\mathbf{x}-\mathbf{x}'|}$, which is as expected because the wave equation in this limit becomes $[\nabla^2+(m^2+\omega^2)/c_S^2]\tilde\alpha\approx-\tilde n/c_S^2$ [see Eq.~\eqref{Green}].

For the remainder of this section, we set $c_S=0$; retaining a nonzero $c_S$ only changes the dispersion relation and the corresponding mode normalization. The radiation field generated by a source with physical period $T\equiv 2\pi/\tilde\Omega$, or auxiliary-time period $\mathtt T\equiv T/(2m_\phi)=2\pi/\Omega$, is given by
\begin{align}
\alpha(\mathtt{t},\mathbf{x}=R\,\mathbf{n})&=\frac{1}{8\pi R}\sum_{n}\frac{n_{(n)}}{k_n^2}\,e^{i\left[\mathbf{k}^{(n)}\cdot\mathbf{x}-\Omega_n \mathtt{t}\right]},
\\
n_{(n)} & = \int d^3x\,e^{-i\mathbf{k}^{(n)}\cdot\mathbf{x}}\left[\frac{1}{\mathtt{T}}\int_0^\mathtt{T}d\mathtt{t}\,n(\mathtt{t},\mathbf{x})\,e^{i\Omega_n \mathtt{t}}\right],
\end{align}
where $\Omega_n\equiv n\Omega$, $\Omega=2m_\phi\tilde\Omega$, and the two-sided expansion uses the signed convention $\mathbf{k}^{(n)}=k_n\mathbf n$, with $k_n=\operatorname{sgn}(n)(\Omega_n^2+m^2)^{1/4}$ for $n\neq0$. Note that $n_0\equiv m/\Omega=\sqrt{4\pi\bar\rho}\,/\tilde\Omega$. We also have
\begin{equation}
\left(\begin{matrix}
\Phi \\ s
\end{matrix}\right)=\frac{1}{8\pi R}\sum_n\left(\begin{matrix}
\kappa_n^\Phi\\ \kappa_n^s
\end{matrix}\right) \frac{n_{(n)}}{k_n^2}\,e^{i\left[\mathbf{k}^{(n)}\cdot\mathbf{x}-\Omega_n \mathtt{t}\right]},
\end{equation}
with
\begin{equation}
\kappa^\Phi_n = \frac{-4\pi\bar\rho}{k_n^2},
\quad
\kappa^s_n = \frac{-i\Omega_n/(2m_\phi)}{k_n^2}.
\end{equation}

Using the far-zone fields above, the time-averaged energy flux is
\begin{equation}\label{quantum_P}
\begin{aligned}
P & = \left\langle R^2\int d\Omega_\mathbf{n}\,\mathbf{S}\cdot \mathbf{n} \right\rangle
=\int d\Omega_\mathbf{n}\sum_{n\ge 1}
\frac{\bar \rho \Omega_n}{128\pi^2m_\phi^3k_n^3}|n_{(n)}|^2
.
\end{aligned}
\end{equation}
The time-averaged energy density is given by
\begin{equation}
\begin{aligned}
\mathcal{E}
&= \left\langle \frac{\bar\rho }{2}|\nabla s|^2 + \bar\rho \Phi\alpha + \frac{\bar\rho }{8m_\phi^2}|\nabla\alpha|^2+\frac{1}{8\pi}|\nabla\Phi|^2 \right\rangle
= \frac{1}{R^2}\sum_{n\ge 1}
\frac{\bar \rho \Omega_n^2}{128\pi^2m_\phi^2k_n^6}|n_{(n)}|^2.
\end{aligned}
\end{equation}
This quantity satisfies $\mathbf{S}\cdot\mathbf n=\mathcal{E}v_g$ mode by mode, where $v_g=k_n^3/(m_\phi\Omega_n)$ is the group velocity of the sound wave. The time-averaged linear momentum density is
\begin{equation}
\begin{aligned}
p_q & = \left\langle -\bar\rho s \partial_q \alpha \right\rangle
=\frac{1}{R^2}\sum_{n\ge 1} \frac{\bar\rho \Omega_n}{64\pi^2m_\phi k_n^5}|n_{(n)}|^2n^q
,
\end{aligned}
\end{equation}
so the time-averaged linear momentum flux is
\begin{equation}\label{quantum_F}
\begin{aligned}
\mathcal{F}_q & = \sum_{n\ge 1}\int d\Omega_\mathbf{n}\,
\frac{\bar\rho}{64\pi^2m_\phi^2 k_n^2}|n_{(n)}|^2n^q
,
\end{aligned}
\end{equation}
which is consistent with the direct evaluation: $\mathcal{F}_q = \left\langle R^2\int d\Omega_\mathbf{n}\,\Pi_q^i n^i \right\rangle$. The time-averaged angular momentum density is
\begin{equation}
\begin{aligned}
j_q & = \left\langle \bar\rho\epsilon_{qkl} x^l s\partial_k \alpha \right\rangle
=\frac{1}{R^2}\sum_n (-i) \frac{\bar\rho \Omega_n}{128\pi^2m_\phi k_n^6}
\epsilon_{qkl} x^l n_{(n)} \partial_k n_{(n)}^*
.
\end{aligned}
\end{equation}
The time-averaged angular momentum flux is
\begin{equation}\label{quantum_tau}
\begin{aligned}
\tau_q & = \operatorname{Re}\int d\Omega_\mathbf n\sum_{n\ge1}(-i)\frac{\bar\rho}{64\pi^2m_\phi^2 k_n^3}\epsilon_{qkl}x^l n_{(n)}\partial_k n_{(n)}^*
,
\end{aligned}
\end{equation}
which is consistent with the direct evaluation: $\tau_q = \left\langle R^2\int d\Omega_\mathbf{n}\,j_q^i n^i \right\rangle$.

\subsection{Common structure of fluxes}\label{sec:common_fluxes}
It is useful to isolate the kinematical structure common to the radiation fluxes produced by a localized periodic source. Let $\omega_n\equiv n\tilde\Omega$, $\rho_{(n)} \equiv \int d^3x\,e^{-ik_n\mathbf{n}\cdot\mathbf{x}}\left[\frac{1}{T}\int_0^Tdt\,\rho_*(t,\mathbf{x})\,e^{i\omega_n t}\right]$. Interestingly, for both the classical fluid and the quantum fluid, the time-averaged energy, linear-momentum, and angular-momentum fluxes can be written in the common form
\begin{equation}\label{common}
\left(\begin{matrix}
P \\ \mathcal{F}_q \\ \tau_q
\end{matrix}\right)
=\sum_{n\ge1}\int d\Omega_\mathbf n
A_n\left(\begin{matrix}
|\rho_{(n)}|^2 \\
\dfrac{k_n}{\omega_n}|\rho_{(n)}|^2 n^q \\
\operatorname{Re}\left[\dfrac{-i}{\omega_n}\epsilon_{qkl}x^l \rho_{(n)}\partial_k \rho_{(n)}^*\right]
\end{matrix}\right),
\end{equation}
where $A_n=\frac{2\bar\rho}{v_g}$ for the classical fluid (with the group velocity $v_g=\frac{c_sk_n}{\Omega_n}$) and the quantum fluid (with $v_g=\frac{k_n^3+\frac{1}{2}c_S^2k_n}{m_\phi\Omega_n}$). The same form also applies to the Klein-Gordon field reviewed in Appendix~\ref{appendix:2}, with $A_n=\frac{\omega_n^2v_g}{8\pi^2}$ and $v_g=\left(1-\frac{n_0^2}{n^2}\right)^{1/2}$ for the radiating harmonics above the mass threshold. This common structure suggests that, once the far-zone amplitude and dispersion relation of the linear, isotropic, nondissipative radiated modes are specified, the conserved charges carried by each mode are fixed kinematically. In principle, Eq.~\eqref{common} can be obtained by directly analyzing the gravitational force exerted on the Newtonian point masses by the perturbed medium.

In the linear-response approximation considered above, $\rho_{(n)}$ is simply fixed by the prescribed periodic source. More generally, if nonlinear medium dynamics modifies the near-zone response while the far-zone radiation remains well described by linear perturbation theory, the time-averaged fluxes retain the same structure as Eq.~\eqref{common}, with the source amplitude $\rho_{(n)}$ replaced by an effective radiative amplitude defined by the asymptotic field, which encodes the nonlinear corrections.

\subsection{Circular orbit}
\begin{figure}[hbt!]
\centering
\includegraphics[width=0.6\textwidth]{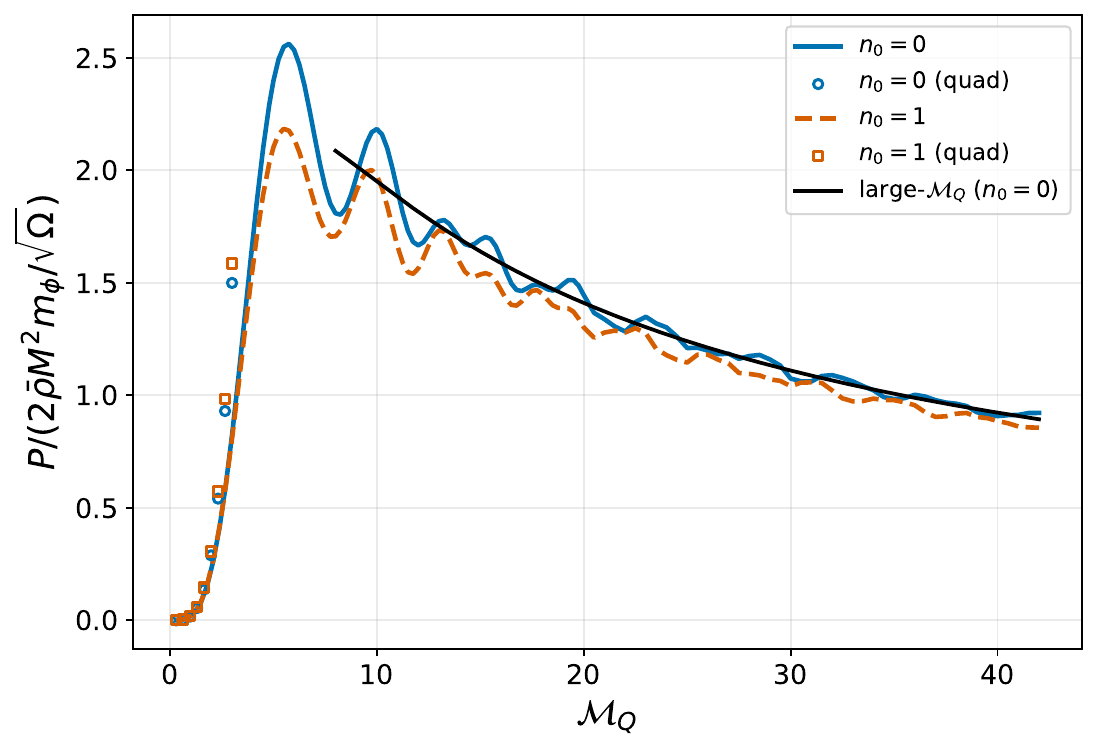}
\caption{The dimensionless circular-orbit function $\mathcal X_Q(\mathcal M_Q)$ for an equal-mass binary, shown for $n_0=0$ and $n_0=1$. The markers show the quadrupole-order approximation in Eq.~\eqref{quantum_circular_quadrupole}, and the black curve gives the large-$\mathcal M_Q$ asymptotic result in Eq.~\eqref{large-M_Q} for $n_0=0$. }\label{fig:quantum_equal_mass_power_curves}
\end{figure}

For the circular binary introduced in Sec.~\ref{sec:Circular orbit}, Eq.~\eqref{quantum_P} gives the energy flux,
\begin{equation}\label{circular_P_2}
P
=\frac{4\pi\bar\rho M^2 m_\phi}{\sqrt{\Omega}}\sum_{n\ge 1}\frac{n I_n}{(n^2+n_0^2)^{3/4}}\equiv \frac{2\bar\rho M^2 m_\phi}{\sqrt{\Omega}}\mathcal{X}_Q
,
\end{equation}
where $I_n$ is given by Eq.~\eqref{I_n}. It is convenient in the quantum-fluid problem to use the dimensionless parameter
\begin{equation}
\mathcal{M}_Q\equiv a\sqrt{\Omega}.
\end{equation}
For an equal-mass circular binary, one has $k_n a=\mathcal{M}_Q(n^2+n_0^2)^{1/4}$, and hence $\mathcal X_Q$ depends on $\mathcal{M}_Q$ and $n_0$; in particular, for $n_0=0$, $k_1a=\mathcal{M}_Q$. Figure~\ref{fig:quantum_equal_mass_power_curves} shows the results for $n_0=0$ and $n_0=1$. We note that a simple estimate suggests the validity condition $m_I m_\phi/|\mathbf v_I|\ll1$ for linear perturbation theory.\footnote{On the de Broglie scale $(m_\phi|\mathbf v_I|)^{-1}$, the ratio of the gravitational potential energy to the kinetic energy is of order $m_I m_\phi/|\mathbf v_I|$.} For a Keplerian binary with $\tilde\Omega^2=M/a^3$ and $q=m_2/m_1\leq1$, requiring this condition for both bodies throughout the orbit gives $\mathcal M_Q\ll\sqrt{2q}[(1-e)/(1+e)]^{1/4}$.

Unlike the classical-fluid flux, the quantum-fluid flux is UV finite because of the $k^4$ dispersion introduced by quantum pressure. To see this, we can repeat the analysis in Sec.~\ref{sec:UV}. In the large-$n$ limit, $k_n/(n\tilde\Omega)\to 2m_\phi/\sqrt{n\Omega}$. For a point source, a real stationary point exists if and only if
\begin{equation}\label{stationary_condition_quantum}
\frac{d}{dt}\left[n\tilde\Omega t-k_n\mathbf n\cdot\mathbf X_I(t)\right]=0
\quad\Longleftrightarrow\quad
\mathbf n\cdot\mathbf v_I(t)\sim n\tilde\Omega/k_n\propto n^{1/2}.
\end{equation}
No finite orbital velocity can satisfy this condition for arbitrarily large $n$.  Hence, the large-$n$ tail has no real stationary point, and the point-source fluxes are UV convergent within the regime of validity of the SP effective theory. We note that the formal limit $n\to\infty$ eventually exceeds the nonrelativistic regime, where the SP description ceases to be applicable. As in the classical-fluid case, a finite-size model such as that considered in Sec.~\ref{uniform-sphere_model} can also be employed here, but we leave this extension for future study.

For an equal-mass circular orbit with $n_0=m/\Omega=0$, asymptotic analysis shows that as $\mathcal{M}_Q\to\infty$
\begin{equation}\label{large-M_Q}
\mathcal{X}_Q \sim 2\pi\frac{2  \ln {\mathcal{M}_Q} -3\ln 2 + \gamma_E}{\mathcal{M}_Q},
\end{equation}
where $\gamma_E$ is the Euler-Mascheroni constant. As $\mathcal M_Q$ increases, the oscillations in $\mathcal X_Q(\mathcal M_Q)$ are damped, while the local mean approaches the asymptotic expression in Eq.~\eqref{large-M_Q}, as shown in Fig.~\ref{fig:quantum_equal_mass_power_curves}. More generally, as $\mathcal{M}_Q\to\infty$, $\mathcal{X}_Q=4\pi\frac{1-3\nu}{\nu}\frac{\ln \mathcal{M}_Q}{\mathcal{M}_Q}+\mathcal{O}(\mathcal{M}_Q^{-1})$.

\subsection[Eccentric orbit]{Eccentric orbit}
\begin{figure}[hbt!]
\centering
\includegraphics[width=\textwidth]{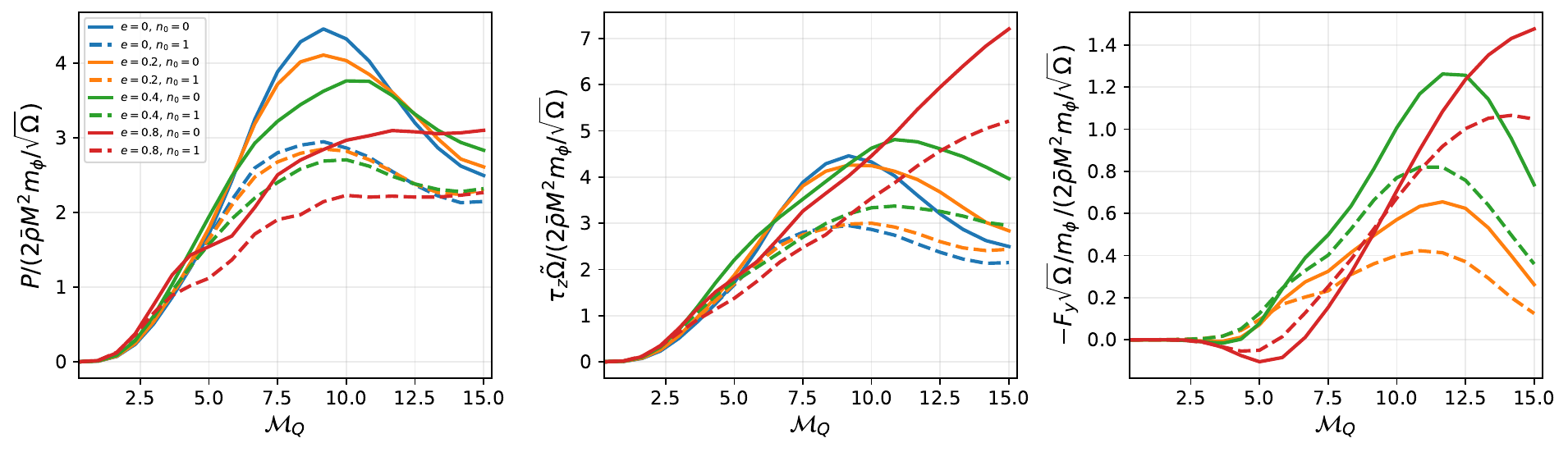}
\caption{Sound-wave fluxes from eccentric binaries with $\nu=0.2$. The panels show $P$, $\tau_z\tilde\Omega$, and $-\mathcal F_y\sqrt{\Omega}/m_\phi$, normalized by $2\bar\rho M^2m_\phi/\sqrt{\Omega}$. These normalized fluxes depend only on $\{e,\mathcal{M}_Q,n_0,\nu\}$.  Solid and dashed curves correspond to $n_0=0$ and $n_0=1$, respectively.}\label{fig:paper_fig3_quantum_nu02_ecc_fluxes}
\end{figure}

For the eccentric binary introduced in Sec.~\ref{sec:Elliptical orbit}, Eqs.~\eqref{quantum_P}, \eqref{quantum_tau}, and~\eqref{quantum_F} give the radiation fluxes:
\begin{align}
P&=\frac{2\bar\rho M^2 m_\phi}{\sqrt{\Omega}}\sum_{n\ge 1} \frac{\int \sin\theta\,d\theta\,d\phi\,|K_n|^2 n}{(n^2+n_0^2)^{3/4}}, \label{quantum_eccentric_P}
\\
\tau_q \tilde \Omega &=
\frac{2\bar\rho M^2 m_\phi}{\sqrt{\Omega}}\text{Re}\sum_{n\ge 1}\frac{(-i)}{(n^2+n_0^2)^{3/4}}\int \sin\theta\,d\theta\,d\phi\,(-K_n\partial_\phi K^*_n)\,\delta_{q,z},
\\
\mathcal{F}_q(\sqrt{\Omega}/m_\phi) &=
\frac{2\bar\rho M^2 m_\phi}{\sqrt{\Omega}} \sum_{n\ge 1} \int \sin\theta\,d\theta\,d\phi\,\frac{2|K_n|^2}{\sqrt{n^2+n_0^2}}\sin\theta\,\sin\phi\,\delta_{q,y}
.
\end{align}
Examples are shown in Fig.~\ref{fig:paper_fig3_quantum_nu02_ecc_fluxes} for $\nu=0.2$; they illustrate the dependence of the fluxes on the eccentricity, the parameter $n_0=\sqrt{4\pi\bar\rho}\,/\tilde\Omega$, and the parameter $\mathcal{M}_Q=a\sqrt{\Omega}$.

As in Sec.~\ref{sec:quadrupole}, the quadrupole-order approximation is valid when $\mathcal{M}_Q=a\sqrt{\Omega}\sim ak_1$ is small. In this approximation, $\boldsymbol{\mathcal{F}=0}$, and
\begin{equation}
\begin{aligned}
P & = \sum_{n\ge 1} \frac{2\pi}{15} \nu^2M^2 \bar\rho \,m_\phi k_n\Omega_n \left(	|\mathcal{I}_{kk}|^2+2 \mathcal{I}_{ij}\mathcal{I}_{ij}^*\right)
,
\\
\boldsymbol{\tau} &= \tau_z\,\mathbf{e}_z =\sum_{n\ge 1} \frac{16\pi}{15} \nu^2M^2\bar\rho \,m_\phi^2 k_n \left(-i
\epsilon_{zkl} \mathcal{I}^*_{ki} \mathcal{I}_{li}
\right)\,\mathbf{e}_z
.
\end{aligned}
\end{equation}
For a circular orbit ($e=0$),
\begin{equation}\label{quantum_circular_quadrupole}
\begin{aligned}
P = \frac{2\pi}{15} \nu^2M^2 \bar\rho m_\phi k_2\Omega a^4
=\tau_z\tilde\Omega
.
\end{aligned}
\end{equation}

\begin{figure}[hbt!]
\centering
\includegraphics[width=0.48\textwidth]{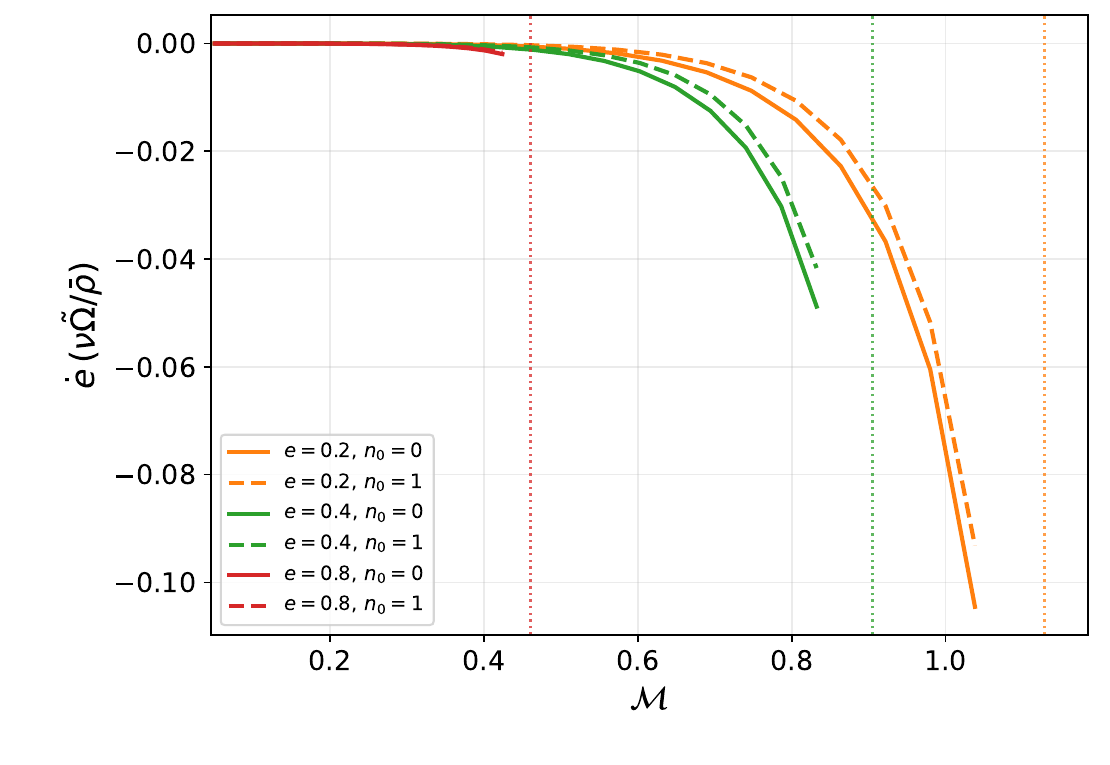}
\includegraphics[width=0.48\textwidth]{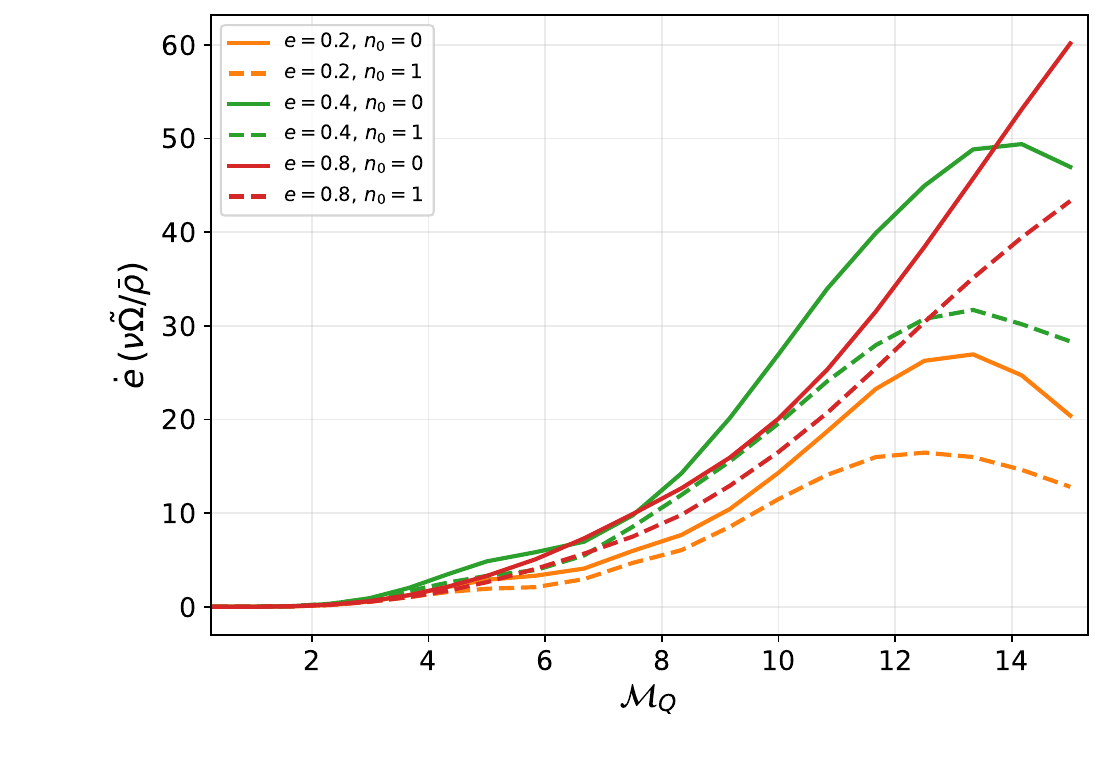}
\caption{Adiabatic eccentricity evolution rate $\dot e$ for eccentric binaries with $\nu=0.2$. The left and right panels show the results for homogeneous classical- and quantum-fluid backgrounds, respectively. Note that $\mathcal{M}=a\tilde\Omega/c_s$, and $\mathcal{M}_Q=a\sqrt{2m_\phi\tilde\Omega}$. The corresponding time-averaged radiation fluxes are shown in Figs.~\ref{fig:paper_fig3_nu02_ecc_fluxes} and~ \ref{fig:paper_fig3_quantum_nu02_ecc_fluxes}, respectively.
}\label{fig:dote}
\end{figure}

The energy and angular-momentum fluxes determine the secular evolution of the osculating Keplerian eccentricity through Eq.~\eqref{orbital_evolution}. Here, we again consider only the contribution from sound radiation. Figure~\ref{fig:dote} compares the resulting $\dot e$ for eccentric binaries in the point-particle theory with the same orbital parameters in classical- and quantum-fluid backgrounds, with $c_S=0$ in the latter case. For the classical-fluid cases shown, $\dot e<0$ throughout the plotted parameter space; see Sec.~\ref{sec:quadrupole} for further discussion. By contrast, for the quantum-fluid cases shown in Fig.~\ref{fig:dote}, the total $\dot e$ remains positive. Interestingly, decomposing the eccentricity evolution rate as $\dot e=\sum_{n=1}^\infty \dot e_n$, with $\dot e_n$ denoting the contribution from the $n$th harmonic, we find that sufficiently high harmonics individually give $\dot e_n<0$ in the quantum-fluid cases. Quantum pressure, however, exponentially suppresses this UV contribution, leaving the full harmonic sum convergent and dominated by the low- and intermediate-order harmonics. For $c_S=0$ and fixed $n_0$, we obtain
\begin{equation}
\dot e =C(e,\nu)\ln \mathcal{M}_Q+\mathcal{O}(1),
\end{equation}
as $\mathcal{M}_Q\to \infty$, with $C(e,\nu)$ given by Eq.~\eqref{C_factor}. It can also be shown for $c_S=0$ that $\dot e>0$ in the quadrupole-order approximation. In our numerical survey within the point-particle theory, we have found no case with $\dot e<0$, although a general analytic proof that $\dot e>0$ remains elusive and is left for future study.

\section{Estimated constraints on the quantum fluid environments of binary pulsars}\label{sec:constraints}

We now give an illustrative estimate using binary-pulsar timing, which has long been used to constrain nonstandard orbital perturbations induced by dark environments~\cite{Iorio:2013pdm,Pani:2015qhr,Caputo:2017dddm,Blas:2019hxz}. The purpose of this section is not to perform a full timing analysis but to translate the absence of excess orbital decay into an order-of-magnitude upper bound on the density of a homogeneous fluid background.

\begin{figure}[hbt!]
\centering
\includegraphics[width=0.55\textwidth]{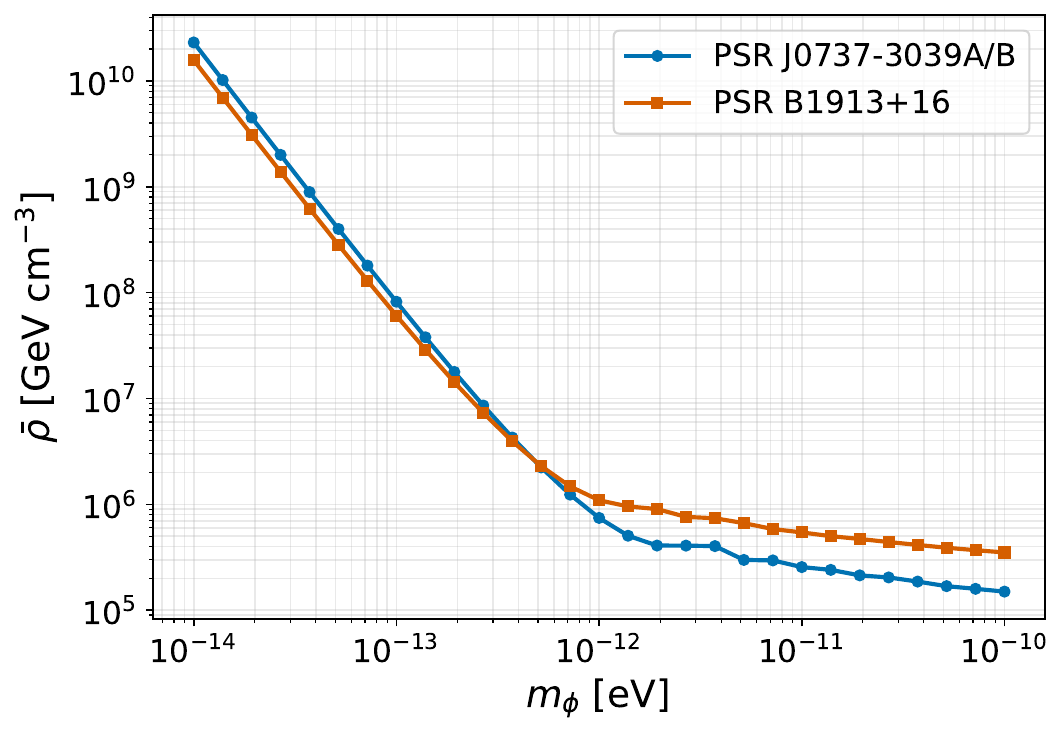}
\caption{Estimated upper bounds on the homogeneous quantum-fluid background density as a function of the scalar mass $m_\phi$, obtained from the orbital-decay budgets in Table~\ref{tab:psr_binary_inputs}.  The two curves correspond to PSR~J0737--3039A/B and PSR~B1913+16. In the parameter space shown, $m_\phi\gg \tilde\Omega$, and the nonrelativistic description of the scalar field is expected to be valid. The domains of validity of the linear perturbation theory, estimated from $\mathcal M_Q\ll\sqrt{2q}[(1-e)/(1+e)]^{1/4}$, are $m_\phi\ll9.2\times10^{-14}\,{\rm eV}$ and $m_\phi\ll3.25\times10^{-14}\,{\rm eV}$ for PSR~J0737--3039A/B and PSR~B1913+16, respectively. Outside this domain, the linear perturbation theory may no longer provide a reliable estimate of the energy flux.}\label{fig:constraint}
\end{figure}

At leading Newtonian order, a binary pulsar is described by an eccentric Keplerian orbit with semimajor axis $a$ and orbital period $T={2\pi}/{\tilde\Omega}$. Using
\begin{equation}
E_{\rm orb}=-\frac{m_1m_2}{2a},
\qquad
\tilde\Omega^2=\frac{M}{a^3},
\end{equation}
the adiabatic energy-balance equation $\dot E_{\rm orb}=-(P_{\rm gw}+P_{\rm med})$ gives
\begin{equation}
\dot T
=
-\frac{6\pi a^{5/2}}{m_1m_2\sqrt{M}}
\left(P_{\rm gw}+P_{\rm med}\right),
\label{Tdot_energy_balance}
\end{equation}
where $P_{\rm med}$ denotes the additional energy flux carried by waves in the surrounding medium. The leading-order energy flux of gravitational radiation, corresponding to 2.5 post-Newtonian (PN) radiation reaction, is~\cite{PhysRev.131.435,PhysRev.136.B1224}
\begin{align}
P_{\rm gw}^\text{2.5PN}
&=
\frac{32}{5}\nu^2M^2a^4\tilde\Omega^6
\frac{
1+\frac{73}{24}e^2+\frac{37}{96}e^4
}{
(1-e^2)^{7/2}
}.
\label{Pgw_constraint}
\end{align}
We specialize to a homogeneous Newtonian quantum-fluid background with $c_S=0$ (as can arise in ultralight scalar dark-matter scenarios~\cite{Hui:2016ltb}) at rest in the binary center-of-mass frame and identify $P_{\rm med}$ with the sound-wave energy flux given by Eq.~\eqref{quantum_eccentric_P}. For the observational comparison, we write
\begin{equation}
\dot T^{\rm th}
=
\dot T^{\rm GR}
+
\dot T^{\rm med},
\qquad
\dot T^{\rm med}
=
-\frac{6\pi a^{5/2}}{m_1m_2\sqrt{M}}P_{\rm med}
\leq 0,
\label{Tdot_medium}
\end{equation}
where $\dot T^{\rm GR}$ is given by Eq.~\eqref{Tdot_energy_balance} with $P_\text{med}=0$. We define $\Delta\dot T \equiv \dot T^{\rm meas}-\dot T^{\rm GR}$. Because an additional dissipative channel makes $\dot T$ more negative, we require the theoretical prediction not to fall below the lower $2\sigma$ observational limit,
\begin{equation}
\dot T^{\rm GR}+\dot T^{\rm med}
\geq
\dot T^{\rm meas}
-
2\sigma_{\dot T^{\rm meas}} .
\end{equation}
Equivalently,
\begin{equation}
-\dot T^{\rm med}
=
\frac{6\pi a^{5/2}}{m_1m_2\sqrt{M}}P_{\rm med}
\leq
B_{2\sigma},
\qquad
B_{2\sigma}
\equiv
\max\!\left(
0,\,
2\sigma_{\dot T^{\rm meas}}
-
\Delta\dot T
\right).
\label{Tdot_budget}
\end{equation}
We adopt the $n_0=0$ approximation when extracting the bounds. Substituting the resulting upper limits on $\bar\rho$ back into $n_0=\sqrt{4\pi\bar\rho}/\tilde\Omega$ confirms that the effect of a nonzero $n_0$ is negligible throughout the parameter range shown. We consider PSR J0737--3039A/B and PSR B1913+16, using the measured binary parameters and orbital-period derivatives listed in Table~\ref{tab:psr_binary_inputs}. The resulting upper limits on $\bar\rho$ are shown in Fig.~\ref{fig:constraint}. In this analysis, we adopt the point-particle approximation and neglect finite-size effects associated with the neutron-star radii and other short-distance physics. For the quantum fluid considered here, the flux is UV finite without introducing an additional cutoff. Similar constraints could also be derived for a classical dark-fluid background. As discussed in Sec.~\ref{sec:1}, however, the sound-wave energy flux in this case is UV divergent in the point-particle theory when the source motion becomes supersonic. In that regime, the resulting constraints are sensitive to the short-distance regularization and are therefore not considered here. Finally, we note that, although a smooth Galactic dark-matter halo would not be expected to yield a density around these systems far above the Solar-neighborhood value $\rho_{\rm DM}\sim0.4\,\mathrm{GeV\,cm^{-3}}$, the latter does not characterize the background around these systems and therefore cannot serve as a direct benchmark for the bounds in Fig.~\ref{fig:constraint}.

\begin{table}[t]
\caption{
Binary parameters and orbital-period measurements used to constrain the homogeneous quantum-fluid density. The directly fitted period derivative is denoted by $\dot T^{\rm obs}$, whereas $\dot T^{\rm meas}$ is the corrected value used in the damping test. For PSR J0737--3039A/B, the spindown mass-loss contribution of pulsar A has been subtracted; for PSR B1913+16 (the Hulse-Taylor binary pulsar), the Galactic correction has been applied. The last two rows give $\Delta\dot T=\dot T^{\rm meas}-\dot T^{\rm GR}$ and the one-sided extra-loss budget defined in Eq.~\eqref{Tdot_budget}. For PSR J0737--3039A/B, we use the 3.5PN result for $\dot T^\text{GR}$ reported in Ref.~\cite{Kramer2021DoublePulsar}; for PSR B1913+16, we use the 2.5PN expression \eqref{Pgw_constraint}, as adopted in Ref.~\cite{WeisbergHuang2016B1913}. Derived quantities are evaluated using the unrounded input values before rounding to the displayed precision.
}
\label{tab:psr_binary_inputs}
\begin{tabular}{lcc}
\hline\hline
Quantity
& PSR J0737--3039A/B
& PSR B1913+16
\\
\hline
Reference
& \cite{Kramer2021DoublePulsar}
& \cite{WeisbergHuang2016B1913}
\\
$m_1\,[M_\odot]$
& 1.338185
& 1.438
\\
$m_2\,[M_\odot]$
& 1.248868
& 1.390
\\
$\nu$
& 0.249702
& 0.249928
\\
$T\,[{\rm d}]$
& 0.1022515592973
& 0.322997448918
\\
$e$
& 0.087777023
& 0.6171340
\\
$\dot T^{\rm obs}\,[10^{-12}]$
& $-1.247920$
& $-2.423$
\\
$\sigma_{\dot T^{\rm obs}}\,[10^{-15}]$
& 0.078
& 1
\\
$\dot T^{\rm meas}\,[10^{-12}]$
& $-1.247782$
& $-2.398$
\\
$\sigma_{\dot T^{\rm meas}}\,[10^{-15}]$
& 0.079
& 4
\\
$\dot T^{\rm GR}\,[10^{-12}]$
& $-1.247827$
& $-2.40263$
\\
$\Delta\dot T\,[10^{-15}]$
& 0.046
& 4.63
\\
$B_{2\sigma}\,[10^{-15}]$
& 0.112
& 3.37
\\
\hline
\end{tabular}
\end{table}

\section{Conclusions}
We have formulated the dissipative effects of dynamical friction on a Keplerian binary in a homogeneous Newtonian fluid as a radiation problem and derived the associated energy, linear-momentum, and angular-momentum fluxes for circular and eccentric binaries. This formulation complements the usual wake-force description by directly capturing the dissipative balance. For a classical inviscid barotropic fluid, the point-source harmonic sums become logarithmically UV-sensitive when at least one binary component becomes supersonic, whereas quantum pressure in a Schr{\"o}dinger-Poisson fluid renders the corresponding fluxes UV convergent within the nonrelativistic regime. The classical- and quantum-fluid fluxes share the same kinematical structure, with the medium dependence entering through the dispersion relation and group velocity. As an illustration, we used binary-pulsar orbital decay measurements to place order-of-magnitude bounds on the densities of homogeneous quantum-fluid backgrounds surrounding these systems.

Several extensions of the present work are worth pursuing. One is to relax the assumption that the medium is at rest in the binary center-of-mass frame. A nonzero center-of-mass velocity $\mathbf{V}$ might activate the $n=0$ mode, for which the resonance condition takes the form $\omega(\mathbf{k})=\mathbf{k}\cdot\mathbf{V}$; this steady-wake sector may introduce infrared sensitivity, requiring a physical large-scale cutoff. As we have seen, sound waves from generic eccentric binaries can carry a nonvanishing time-averaged linear-momentum flux. In a self-consistent treatment, the corresponding recoil would cause the binary center of mass to acquire a nonzero velocity relative to the medium, an effect neglected here under the prescribed-orbit approximation. The radiation formulation could also be extended to the relativistic regime~\cite{Barausse:2007df} or to other media, including magnetized fluids~\cite{SanchezSalcedo:2012magnetic,Shadmehri:2012magnetic}, fluids in modified-gravity theories such as modified Newtonian dynamics~\cite{SanchezSalcedo:2009mond}, viscous media~\cite{Katz:2019df} (for a simple viscous barotropic Newtonian fluid, the sound-wave equation becomes $\ddot\alpha-c_s^2\nabla^2\alpha-\nu_L\nabla^2\dot\alpha=4\pi(\bar\rho\,\alpha+\rho_*)$, with $\nu_L>0$), and collisionless Vlasov-Poisson media~\cite{TremaineWeinberg:1984df}, where the response to a periodic source is not captured by a simple set of discrete on-shell radiative modes but involves phase-space resonances. It would also be useful to go beyond the dissipative balance considered here by determining the conservative wake force, which can, for example, induce orbital precession~\cite{2024ApJ...974..216O}, and by incorporating nonlinear corrections to the medium response~\cite{Kim:2010nonlinear}, which are also necessary for reliably estimating the constraints at large scalar-boson masses in Fig.~\ref{fig:constraint}. Finally, for sufficiently large-scale binary systems, such as galaxy binaries~\cite{2023ApJ...953L...2B}, the radiation formulation could be extended to an expanding Friedmann--Lema\^{i}tre--Robertson--Walker background, where the cosmological evolution of the dark-matter medium and the spacetime geometry modify the propagation of the induced perturbations and their backreaction on the source. It remains to ask what the sound of binaries can tell us.

\section*{Acknowledgments}
We thank the anonymous referee for insightful comments and valuable suggestions that improved the manuscript.


\appendix
\section[Time-domain derivation of the frequency-domain waveform]{Time-domain derivation of the frequency-domain waveform}\label{appendix:1}
Here, we provide an explicit derivation of the far-zone waveform using the time-domain retarded Green's function for a real massive Klein-Gordon field in flat spacetime. The retarded Green's function satisfying $(\square_x-m_\phi^2)G(x,x')=-\delta^4(x-x')$ with $\square=-\partial_t^2+\nabla^2$ and $m_\phi^2>0$ is~\cite{Alsing_2012}
\begin{equation}
G(x,x')=\frac{\delta(t'-(t-|\mathbf{x}-\mathbf{x}'|) )}{4\pi |\mathbf{x}-\mathbf{x}'|}
-\Theta( (t-|\mathbf{x}-\mathbf{x}'|)-t')\frac{m_\phi J_1(m_\phi\sqrt{-(x-x')^2})}{4\pi \sqrt{-(x-x')^2}}
,
\end{equation}
where the tail term is supported inside the past light cone, $t-t'>|\mathbf{x-x'}|$. The solution of $(\square-m_\phi^2) \phi =-n(t,\mathbf{x})$ is given by
\begin{equation}
\begin{aligned}
\phi(t,\mathbf{x})=&\int d^3x'\,\frac{n(t-|\mathbf{x}-\mathbf{x}'|,\mathbf{x}')}{4\pi|\mathbf{x}-\mathbf{x}'|}
\\
&-\int d^3x'\int_{-\infty}^{t-|\mathbf{x}-\mathbf{x}'|} dt'\,\frac{m_\phi J_1\left(m_\phi\sqrt{(t-t')^2-|\mathbf{x}-\mathbf{x}'|^2}\right)\,n(t',\mathbf{x}')}{4\pi \sqrt{(t-t')^2-|\mathbf{x}-\mathbf{x}'|^2}}
.\label{time_domain_solution}
\end{aligned}
\end{equation}
The source can be written as
\begin{equation}
n(t',\mathbf{x}')=\int_{-\infty}^\infty \frac{d\omega}{2\pi}\, n(\omega,\mathbf{x}')\,e^{-i\omega t'}.
\end{equation}
If $R=|\mathbf{x}|$ is sufficiently large, we have
\begin{equation}
\begin{aligned}
&-\frac{m_\phi }{4\pi}\int_{-\infty}^{t-r} dt' \frac{J_1\left(m_\phi\sqrt{(t-t')^2-r^2}\right)}{\sqrt{(t-t')^2-r^2}}e^{-i \omega t'}
\\
&=
-\frac{1}{4\pi r}\left\{\int_0^\infty dz \frac{J_1\left(z\right)\,e^{i \omega r\sqrt{1+[z/(m_\phi r)]^2}}}{\sqrt{1+[z/(m_\phi r)]^2}}\right\}e^{-i\omega t}
\\
&
\approx -\frac{1}{4\pi r}\left(e^{i\omega r}-e^{i r\,\operatorname{sgn}(\omega)\sqrt{\omega^2-m_\phi^2}}\right)e^{-i\omega t}
,
\end{aligned}
\end{equation}
where $r=|\mathbf{x}-\mathbf{x}'|$. Here, we used the result~\cite{Alsing_2012},
\begin{equation}\label{useful_result}
\lim_{r\to\infty}\int_0^\infty dz \frac{J_1\left(z\right)\,e^{i \omega r\sqrt{1+[z/(m_\phi r)]^2}}}{\left\{1+[z/(m_\phi r)]^2\right\}^{n/2}}
\to e^{i\omega r} - \left[1-\left(\frac{m_\phi}{\omega}\right)^2\right]^{\frac{n-1}{2}}e^{i r\,\operatorname{sgn}(\omega)\sqrt{\omega^2-m_\phi^2}}
,
\end{equation}
for $\omega^2>m_\phi^2$. Equation~\eqref{time_domain_solution} can then be written as
\begin{equation}
\begin{aligned}
\phi(t,\mathbf{x})&\approx\int_{-\infty}^\infty \frac{d\omega}{2\pi}e^{-i\omega t}\int d^3x'\,n(\omega,\mathbf{x}')\left\{\frac{e^{i\omega r}}{4\pi r}-\frac{1}{4\pi r}\left[ e^{i\omega r}-e^{i r\,\operatorname{sgn}(\omega)\sqrt{\omega^2-m_\phi^2}} \right]\right\}
\\
&=\int_{-\infty}^\infty \frac{d\omega}{2\pi}\int d^3x'\,n(\omega,\mathbf{x}')\frac{e^{i \left[r\,\operatorname{sgn}(\omega)\sqrt{\omega^2-m_\phi^2}\,-\,\omega t\right]}}{4\pi r}
,
\end{aligned}
\end{equation}
where the $\omega$ integral is restricted to $\omega^2>m_\phi^2$. For $\mathbf{x}=R\,\mathbf{n}$ with $|\mathbf{n}|=1$, using $r\approx R-\mathbf{n}\cdot\mathbf{x}'$, the radiation field can be written as
\begin{equation}\label{result_1}
\phi(t,\mathbf{x})\approx \frac{1}{4\pi R} \int_{-\infty}^\infty \frac{d\omega}{2\pi}\left[\int d^3x'\,e^{-i\mathbf{k}_\omega\cdot\mathbf{x}'}n(\omega,\mathbf{x}')\right]e^{i (\mathbf{k}_\omega\cdot\mathbf{x}-\omega t)},
\end{equation}
with $\mathbf{k}_\omega=\operatorname{sgn}(\omega)\sqrt{\omega^2-m_\phi^2}\,\mathbf{n}$. For a tachyonic mass term, $m_\phi^2=-m^2<0$ with $m>0$, the retarded Green's function follows by the analytic continuation $m_\phi\to i m$. For the real-frequency radiation, this gives $\mathbf{k}_\omega=\operatorname{sgn}(\omega)\sqrt{\omega^2+m^2}\,\mathbf{n}$.

\section[Radiation of a Klein-Gordon field]{Radiation of a Klein-Gordon field}\label{appendix:2}
We consider a real massive Klein-Gordon field $\phi(t,\mathbf{x})$ coupled to an external source $\rho(t,\mathbf{x})$ in flat spacetime. With the mostly plus metric signature, the Lagrangian is $\mathcal{L}=-\frac{1}{2}(\partial\phi)^2-\frac{1}{2}m_\phi^2\phi^2+\rho\phi$, and the wave equation reads $(\partial_t^2-\nabla^2+m_\phi^2)\phi=\rho$. For a localized periodic source with $\rho(t,\mathbf{x})=\rho(t+T,\mathbf{x})$, the radiation field is given by
\begin{align}
\phi(t,\mathbf{x}=R\,\mathbf{n})&=\frac{1}{4\pi R}\sum_{|n|\ge n_0}\rho_{(n)}\,e^{i\left[\mathbf{k}^{(n)}\cdot\mathbf{x}-\Omega_nt\right]},
\\
\rho_{(n)} & = \int d^3x\,e^{-i\mathbf{k}^{(n)}\cdot\mathbf{x}}\left[\frac{1}{T}\int_0^Tdt\,\rho(t,\mathbf{x})\,e^{i\Omega_n t}\right],
\end{align}
where $n_0=m_\phi/\Omega$, $\Omega=2\pi/T$, and $\mathbf k^{(n)}=k_n\mathbf n$, with $k_n=\operatorname{sgn}(n)\sqrt{\Omega_n^2-m_\phi^2}$. The corresponding time-averaged radiation fluxes are~\cite{Chen:2025mwl}
\begin{equation}
\begin{aligned}
P & =
\sum_{n\ge n_0}\frac{\Omega_n^2}{8\pi^2}\left(1-\frac{n_0^2}{n^2}\right)^{1/2}\int d\Omega_\mathbf{n}  \left| \rho_{(n)}\right|^2
,
\\
\mathcal{F}_q & =
\sum_{n\ge n_0}\frac{\Omega_n^2}{8\pi^2}\left(1-\frac{n_0^2}{n^2}\right) \int d\Omega_\mathbf{n}\, |\rho_{(n)}|^2 n^q,
\\
\tau_q&=\operatorname{Re}\sum_{n\ge n_0}(-i)\frac{k_n}{8\pi^2}\int d\Omega_\mathbf n\,\epsilon_{qkl}x^l \rho_{(n)}\partial_k \rho_{(n)}^*.
\end{aligned}
\end{equation}

\clearpage
\bibliographystyle{apsrev4-2}
\bibliography{paper}

\end{document}